\documentclass[12pt,onecolumn,draftcls]{IEEEtran}
\usepackage{color}
\usepackage{listings}
\usepackage{amssymb}
\usepackage{amsmath}
\usepackage{amsfonts}
\usepackage{graphicx}
\usepackage{epsf, epsfig, textcomp}
\usepackage{subfigure}
\usepackage{float}
\usepackage{dsfont}
\usepackage{balance}
\usepackage{amsthm}
\usepackage{array}

\usepackage{epstopdf}

\newcommand{\HRule}{\rule{\linewidth}{0.5mm}}
\usepackage{cite}
\usepackage{url}
\usepackage{enumerate}

\newtheorem{theor}{Theorem}
\newtheorem{lem}{Lemma}

\newtheorem{prop}{Proposition}

\makeatletter
\def\footnoterule{\relax%
  \kern-5pt
  \hbox to \columnwidth{\hfill\vrule width \columnwidth height 0.4pt\hfill}
  \kern4.6pt}
\makeatother

\begin{document}
%
\title{Inference-Based Distributed Channel Allocation \\in Wireless Sensor Networks}
%
%
%

\author{Panos~N.~Alevizos,~\IEEEmembership{Student Member,~IEEE},
        Efthymios~A.~Vlachos,
        and Aggelos~Bletsas,~\IEEEmembership{Senior Member,~IEEE}
%
\vspace{-0.1 in}
\thanks{Part of this work  is presented at IEEE Global Communications Conference (GLOBECOM) 2014, Austin, Texas, USA.
This work was supported by the ERC-04-BLASE project, executed in the context of the
"Education \& Lifelong Learning" Operational Program of the National
Strategic Reference Framework (NSRF), General Secretariat for
Research \& Technology (GSRT), funded through European
Union-European Social Fund and Greek national funds. 

The authors are
with   School of Electrical and Computer Engineering (ECE), Technical University of Crete, Chania 73100, Greece
(e-mail: palevizos@isc.tuc.gr, efthimios.vlachos@gmail.com, and aggelos@telecom.tuc.gr). 
}}


\maketitle

\begin{abstract}
Interference-aware resource allocation of time slots and frequency channels in single-antenna, half-duplex radio wireless sensor networks (WSN) is challenging. 
Devising distributed algorithms for such task further complicates the problem. This work studies WSN joint time and frequency channel allocation for a given
routing tree, such that: a) allocation is performed   in a 
fully {distributed} way, i.e., information exchange is only performed among neighboring WSN terminals, within communication up to two hops, and
b) detection of potential interfering terminals is simplified and can be practically realized.
 The algorithm imprints space, time, frequency and radio hardware constraints into a loopy factor graph and performs iterative
message passing/ loopy belief propagation (BP) with randomized initial priors. 

Sufficient conditions for  convergence to a valid solution are offered, for the first time in the literature, exploiting the structure
 of the proposed factor graph. Based on theoretical findings, modifications of BP are devised that i) accelerate convergence to a valid solution and ii) reduce computation cost.
Simulations reveal promising throughput results of the proposed distributed algorithm,  
  even though it utilizes simplified interfering terminals set detection. Future work could modify the constraints such that other disruptive wireless technologies (e.g., full-duplex radios or network coding) could be accommodated within the same inference framework. 
\end{abstract}



%
\IEEEpeerreviewmaketitle

\section{Introduction}
\label{sec:intro}

Designing efficient channel allocation algorithms, i.e., assigning time slots and/or frequency channels in resource constrained wireless sensor networks (WSNs), may offer tremendous interference mitigation opportunities   and subsequent throughput,  delay, or energy-efficiency  improvements 
\cite{ChWaCh:06, YaStTiJLin:08, CheYuChFaSh:11,
SaiXuLuChen:11,  ChChAgZe:05, ZoHuYaHeStAb:06,  LeHenAb:08, WaWaFuGuN:09,
 ChOcLuChSt:06, ZhTiSta:05, KimShiCha:08, TaSGuB:11}.
WSNs support a wide 
range of applications, including environmental sensing, smart buildings, medical care, 
micro-climate monitoring and plethora of other industry and military applications.
WSNs differ from traditional wireless ad-hoc or heterogeneous (5G) networks in the
following aspects:
(a) each WSN terminal is low-cost, low-power, single-antenna with half-duplex radio,
(b) the number of  available frequency channels in current  WSNs  may be limited in practice,
(c) the available bandwidth of WSN terminals may be also limited, e.g.,   $250$ kbps in
802.15.4 networks,
(d) memory  and processing power are typically limited per WSN terminal, e.g., $10$ kByte memory 
and $8$ MHz MSP430 microcontroller in TelosB motes \cite{TelosB}, 
and  (e) the packet payload may be small to minimize delay and power consumption. 

The problem of channel allocation becomes even more challenging in large-scale WSNs, where the computational burden   should  be dispensed across all terminals,
pointing towards \emph{distributed} protocols \cite{CheYuChFaSh:11,
 SaiXuLuChen:11, LinRas:09, InHoJan:11, ChChAgZe:05, HaKuMaPaSr:09,
VedKakLaSi:06}. Centralized protocols may be prohibitive for large-scale WSNs with resource constrained terminals due to computation cost, as well as large delays at WSN terminals in the vicinity of the central processing unit.  On the other hand, a distributed protocol requires the following: (a) local knowledge at each WSN terminal,
 e.g., that are its interferers \cite{CheYuChFaSh:11} or its up to two-hop neighbors   in  the routing tree  \cite{ZoHuYaHeStAb:06},   and (b) 
a message-passing (MP) communication mechanism among neighboring terminals, based on specific
 synchronous or asynchronous schedule \cite{BertTsi:1989}. 
 
 Distributed WSN frequency channel allocation algorithms are presented in
\cite{CheYuChFaSh:11} and \cite{SaiXuLuChen:11}; in\cite{CheYuChFaSh:11}, a
game theory-based algorithm is employed in order to minimize the    total    number of interfering links,
 while in \cite{SaiXuLuChen:11},  a  distributed algorithm  is proposed,
which eliminates the remaining interference links in the WSN, by constructing
a conflict-free TDMA schedule. 
In addition, works in \cite{CheYuChFaSh:11} and \cite{SaiXuLuChen:11} 
make the implicit assumption that interference connectivity among all WSN terminals is precisely known.
Interference connectivity of a WSN terminal is defined as the set of terminals that interfere the transmission
 or the reception of that terminal (depending  on the utilized interference set detection protocol).
In many cases, the signal-to-noise ratio (SNR) at a receiving terminal $j$ may be degraded by
\emph{simultaneous} transmissions from a set $\mathcal{I}$ of several WSN terminals, whose \emph{individual} transmission may not degrade significantly the SNR at terminal $j$;
in that case, terminals in set $\mathcal{I}$ cannot be easily identified and incorporated in the interference connectivity set of terminal $j$. This is another reason why prior art has introduced the notion of \emph{interference radius}, as opposed to communication radius \cite{YaStTiJLin:08}.


Work in \cite{ChWaCh:06} calculates a TDMA schedule for packet radio networks,
assuming single-frequency channel radio terminals. More specifically, constraints based on
link connectivity up to two hops are encoded using a factor graph (FG), assuming that simultaneous transmission from two (or more) neighboring terminals is always harmful. Thereinafter, the loopy belief 
propagation (BP) runs between neighboring terminals in order to find out a global time-slot schedule that
adheres to all (local) constraints. 

Due to the loopy nature of the proposed FG, the mathematical toolbox to guarantee convergence to a valid solution,
or even convergence to a fixed point (that may not be a  valid solution) is restricted.
For the latter case, only a few exemplary methodologies and results exist in
 the literature   \cite{WeFr:01, TaJo:02,  Hes:04, BrMeZe:05,  IhFiWill:05, MoKap:07,  BaShSh:08, BaBoChZe:08,  BaMo:11,  NoorWein:13}.
 In a general loopy probabilistic graphical model (PGM), where BP is executed, convergence to a fixed point does not necessarily imply correctness, i.e., convergence
to a valid (or correct) solution, apart from special cases, as in Gaussian BP \cite{WeFr:01}  or maximum weight matching problems \cite{BaShSh:08, BaBoChZe:08};
 convergence to a correct solution is a critical part in our channel allocation challenge, crafted as a feasibility problem. 

\begin{figure}[!t]
    \centering
        \includegraphics[scale=0.5]{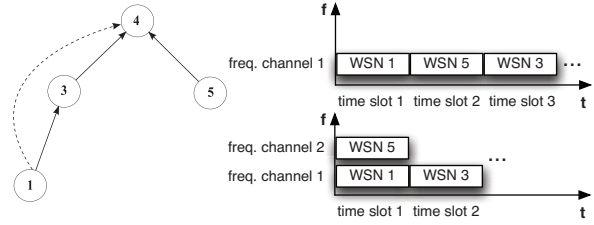}
    \caption{Transmissions of WSN terminals under two channel allocation schemes, for the the specific routing tree (solid lines) and specific interference link (dotted line). Top allocation is based on time slots, bottom allocation is based on both time slots and frequency channels. This work offers such allocations with distributed inference, under convergence, correctness and computation cost guarantees.}
    \label{fig:dof}
\end{figure}

On the other hand, joint time and frequency channel allocation accounts not only for time but also for frequency channelization,
which is an essential part of contemporary multi-channel radio modules, 
as it provides additional degrees of freedom and thus, potential for more efficient networking. For example, consider the $4$-terminal network of Fig.~\ref{fig:dof} with the specific routing tree topology (solid lines) and two channel allocation schemes, one with time slots (top) and a second with both time and frequency channels (bottom); under single-frequency half-duplex radios, $3$ time slots are needed so that information from the leaf terminals reaches sink terminal $4$ (top allocation); that is due to the fact that transmission of terminal $1$ towards parent terminal $3$ is interfering receiver $4$ (and such interfering link between $1$ and $4$ is depicted  as a dotted line in Fig.~\ref{fig:dof}). With multiple frequency channels, the required time slots are reduced to $2$ (bottom allocation), even with half-duplex radios, offering smaller delay and higher effective throughput, at the expense of additional bandwidth.

 From an implementation point of view,  identification of potential interferers, i.e.,
interference set detection,  is a prerequisite step for any joint time-frequency allocation algorithm. 
In addition to the above,  single-antenna, half-duplex radios impose extra hardware
constraints that have to be taken into account, rendering time slot and frequency channel 
allocation a challenging task for WSNs.

This work extends time slot allocation in
 \cite{ChWaCh:06} and addresses distributed, joint time slot and frequency 
channel allocation.  Following the RID framework in \cite{ZhTiSta:05},  practical, low-complexity
   interference set detection is utilized,  based on signal-to-noise-and-interference ratio (SINR). 
In addition,  a routing tree is assumed,  as in well-known WSN protocols \cite{ChTa:00, AnZeChSa:08, DeKoStDe:11},
with sink as the tree root. 
The proposed algorithm is a modified version of loopy BP  running
 on a carefully crafted FG that encodes both time and frequency-based constraints,
 taking into account   routing and interference connectivity, as well as radio hardware constraints, e.g., due to half-duplex operation 
 or the fact that each radio can tune at a single frequency channel at a time.
 Each WSN terminal 
is associated with  specific variables and factor nodes of the FG, so that message passing (MP) 
   with  neighboring  WSN terminals in communication connectivity   is only needed.  
The MP schedule of the proposed FG requires
 the transmission of a single real number per directed FG edge and can be implemented in a distributed manner.
  The objective  is to find a feasible frequency-time allocation
 that adheres to specific communication, routing, interference and radio hardware constraints. 

The contributions of this work are summarized below:
\begin{enumerate}[A.]
\item A joint time slot and frequency channel allocation algorithm is offered, based on  
loopy BP; the algorithm is distributed, 
since each WSN terminal needs to communicate  with up to two-hop neighboring WSN terminals 
in communication radius (i.e., communication connectivity), associated with routing and interference.  
Interference set detection is practical and based on sensitivity and  SINR  at each receiving radio terminal.  

\item Sufficient conditions for convergence to correct solution are offered   for  loopy BP, for the first time in the literature (to the best of our knowledge),
 exploiting the structure of the underlying PGM; the latter is crafted under the specific problem constraints.

\item Computation cost reduction methods are offered, based on precomputed feasibility sets found with binary search;  such methodology is important since the complexity of the
 underlying loopy BP algorithm is exponential in the PGM degree, which in turn depends on network connectivity. 

\item An interesting tradeoff is offered between remaining interference of the offered solution and computation time. Furthermore, random local re-initialization among WSN terminals running the algorithm is introduced, showing significant convergence acceleration. 
\end{enumerate}

The inherent expressive power of MP/BP inference framework, including asynchronous scheduling capabilities, could spark interest for distributed solutions in other network scenarios, offering perhaps a new, fresh look at an old networking problem \cite{GeNeeTa:06, LiEf:07}. Compared to conference version \cite{AlVlBl:14}, this work provides
a detailed exposition of the adopted interference set detection procedure,
offers new sufficient convergence conditions on exact solution (with proof),  examines complexity
issues and proposes acceleration techniques; in addition, numerical results study  large-scale
WSN topologies and quantify  how quickly the proposed algorithm   converges to an exact solution
under fully distributed operation using the proposed modification of BP.

%

\emph{Notation}:
Symbol $\mathcal{U}[0,1]$ denotes the continuous uniform distribution over the
(closed) interval $[0,1]$. 
Symbols $\mathds{B}$ and $\mathds{N} \triangleq \{1,2,\ldots\}$ denote the set of binary and natural numbers, respectively. 
The operator $|\cdot|$ stands for the cardinality of a set, i.e., $|\mathds{B}|=2$.
The   number of the non-zero elements of a vector $\mathbf{x}$, is denoted
$ \|\mathbf{x} \|_0$. 
The vector  comprised of variables associated with  an arbitrary index set  $\mathcal{A}$, is denoted as
 $\mathbf{x}_{\mathcal{A}} =\left\{ x_{y} \right\}_{y \in \mathcal{A}}$.
Symbol $\mathsf{1}\{ \cdot\}$ stands for
the indicator function that returns one if the statement within the brackets is true, and zero, otherwise.

%

\section{Problem Formulation and Interferer Set Detection}
\label{sec:sys}
\label{sec:sys_mod}
 A WSN consisting of $N$ single-antenna, half-duplex radio   terminals is considered. Sink terminal operates in receiver mode only. A terminal needs a time slot to transmit a packet
and each transmission frame consists of $M$ equal-length time slots. The available frequency
bandwidth is divided into $K$ orthogonal frequency channels.  Let $\mathcal{M} \triangleq \{1,2,\ldots, M\}$
and $\mathcal{K} \triangleq \{1 ,2 \ldots, K\}$ be  the set of available time slots and frequency channels, respectively.
Let $\mathcal{N}$  be the set of all terminals in the WSN (sink included); a communication link $(i,j)$
between two terminals $i,j \in \mathcal{N}$ exists  if during $i$'s transmission,  the received signal strength at terminal $j$
 is above its receiver sensitivity. 

\begin{table}[!t]
\renewcommand{\arraystretch}{1.0}
\centering
\caption{Notation and Relations in WSN Routing Tree}
\begin{tabular}{|c|c|}
\hline
{\bf Symbol } & {\bf Relations in WSN routing tree connectivity}\\
 \hline
\rule{0pt}{8pt}
$\mathcal{N}$ &  all  WSN terminals, i.e., $\mathcal{N } \triangleq\{ 1,\ldots,N\}$\\
\hline
$\mathtt{s}$ & the sink terminal, $\mathtt{s} \in \mathcal{N }$\\
\hline
$\mathcal{N}_{\backslash \mathtt{s}}$ &  all  WSN terminals except sink,  i.e., $\mathcal{N }_{\backslash \mathtt{s}} \triangleq \mathcal{N } \backslash \mathtt{s} $\\
\hline
\rule{0pt}{8pt}
${\rm par}(i)$  & the (unique) parent  of terminal $i$\\
\hline
\rule{0pt}{8pt}
$\mathcal{N}_{\rm ch}(i)$ & children of terminal $i$: $\mathcal{N}_{\rm ch}(i) \hspace{-0.022 in}\triangleq
 \hspace{-0.022 in}\left\{ i' \in \mathcal{N}  \hspace{-0.01 in} :  \hspace{-0.01 in}{\rm par}\!\left(i' \right) = i
\right\}$ \\
\hline
\rule{0pt}{8pt}
 $\mathcal{N}_{\rm sib}(i)$ & the set of sibling terminals of  terminal $i$,  i.e.,   \\
& $\mathcal{N}_{\rm sib}(i) \triangleq  \left\{ i' \in \mathcal{N}_{\backslash \mathtt{s}} \backslash  i : i' \in
\mathcal{N}_{\rm ch}\!\left({\rm par}\!\left(i' \right) \right) \right\}$ \\
\hline
\rule{0pt}{8pt}
$\mathcal{N}_{\rm OneH}(i)$ & the set of one-hop neighbors of  terminal $i$,  i.e., \\
& $\mathcal{N}_{\rm OneH}(i) \triangleq \big\{ i' \in \mathcal{N}_{\backslash \mathtt{s}} : i'  \in
\mathcal{N}_{\rm ch}(i) \cup   i ~  \cup $
$ {\rm par}(i) \big \}$ \\
\hline
\rule{0pt}{8pt}
 & the set of two-hop neighbors of  terminal $i$,  i.e., \\
$\mathcal{N}_{\rm TwoH}(i)$ & \hspace{-0.2 cm}$\mathcal{N}_{\rm TwoH}(i) \!\triangleq\! \bigg\{ i' \in  \mathcal{N}_{\backslash \mathtt{s}}  \! :\! i' \in  \mathcal{N}_{\rm OneH} ({\rm par}(i)) \cup$ \\
& $\left ( \bigcup_{j \in \mathcal{N}_{\rm ch}(i) } \mathcal{N}_{\rm OneH}(j) \right) \bigg\}$ \\
\hline
 \end{tabular}
\label{notation_table}
\end{table}

A tree routing connectivity  is assumed  \cite{GonDaMiYu:11}, abbreviated as $\mathcal{T} = (\mathcal{N},\mathcal{C}_{\mathcal{T}})$, 
where $ \mathcal{C}_{\mathcal{T}}$  is the set of the edges of the WSN routing  tree after the execution of routing algorithm. 
Table \ref{notation_table} summarizes the adopted notation related to the routing connectivity (for exposition purposes
 the defined sets exclude sink terminal $\mathtt{s}$).
 
For any routing  link $(i,j) \in \mathcal{C}_{\mathcal{T}}$,  (i.e., $j = {\rm par}(i)$),  
the set of potential interferers of  link $(i,j)$ consists of any subset $\mathcal{I}$ of  terminals that
 degrade the signal-to-noise-ratio  (SNR) at receiver $j$; in other words,   WSN terminals in $\mathcal{I}$  satisfying
\begin{equation}
{\rm SINR}_{i \rightarrow j}^{\mathcal{I}} =
\frac{P_{i}|{h}_{i,j}|^2}{\sigma_j^2+ \sum_{i' \in \mathcal{I}} P_{i' }|{h}_{i' ,j}|^2} < \theta,
\label{eq:SINR_all}
\end{equation}
should be included in the  set of links' $(i,j)$ potential interferers.
In Eq.~\eqref{eq:SINR_all}, $P_{i}$ is the power of transmitter $i$, ${h}_{i,j}$
is the instantaneous channel gain coefficient between transmitter $i$ and
receiver $j$ incorporating both large and small scale  fading,
  $\sigma_j^2$ is the thermal noise power at receiver $j$, 
and  $\theta$ is a threshold parameter that
depends on the receiver sensitivity.\footnote{Typical values around $-95$dBm are found in
the literature and depend on receiver noise figure, transmission
bandwidth, receiver temperature and required SNR. Numerical
results assume receiver sensitivity at $-100$dBm.} 
Executing the SINR test in Eq.~\eqref{eq:SINR_all}, 
   requires a search on 
all possible subsets of interfering terminals, which is prohibitive for resource constrained WSN terminals.
More importantly, discovery of interfering terminals may be impossible, since a WSN terminal may contribute 
to the sum of the denominator in Eq.~\eqref{eq:SINR_all}, but with power which may not be adequate for receiver $j$ to properly decode a packet and discover the identity of the interferer;  
the superposition of several undecodable signals can contribute to the the sum in the denominator of Eq.~\eqref{eq:SINR_all}.

Even though modeling of interference in this work adheres to Eq.~\eqref{eq:SINR_all}
(utilized also during numerical results), discovery of interfering terminals adopts a 
modified version of lightweight RID protocol \cite{ZhTiSta:05}, simplifying interference set identification. 
   Let  us denote $\mathcal{N}_{\rm pint}(j) $  the set of potential interferers of terminal $j$, including all WSN terminals $i'$ satisfying the following conditions:  a) link between $i'$ and $j$ does not belong to the routing tree
(i.e., $(i',j) \notin \mathcal{C}_{\mathcal{T}}$) and b)
 reception of $i'$ transmission at $j$ is above $j$'s receiver sensitivity.
 Terminal $i' \in \mathcal{N}_{\rm pint}(j)$ is an actual  interferer of transmission from child $i$ to parent $j$
 (or simply interferer of $j$), if the following condition holds
\begin{equation}
{\rm SINR}_{i \rightarrow j}^{i'} =
\frac{P_{i}|{h}_{i,j}|^2}{\sigma_j^2 + P_{i'}|{h}_{i',j}|^2} < \theta.
\label{eq:SINR_one}
\end{equation}
Link between terminal $i'$ and $j$ is an \emph{interfering link}.
 Discovery of interferes for a specific link (or equivalently for a specific receiver) requires examination of the above test for   all terminals $i' \in \mathcal{N}_{\rm pint}(j)$.
 Examination of the above test requires linear complexity on the number of potential interferers.
 This simplification, even though underestimates the number of potential interferers,
reduces the required overhead needed for interfering set identification. 
Moreover, the above test can be practically applied among WSN terminals  neighboring to $j$, that can be properly decoded and identified by $j$.
 
\begin{figure}[!t]
    \centering
        \includegraphics[scale=0.4]{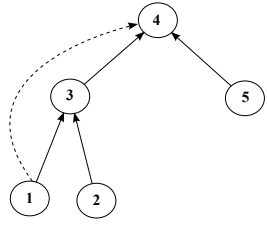}
    \caption{A simple WSN routing tree with $5$ sensor terminals. Dashed and solid lines depict
    interference and routing connectivity, respectively. Terminal 1 interferes reception at terminal 4 (or transmission of terminal 5), even though it transmits to terminal 3 only.}
    \label{fig:simple_net}
\end{figure}

Let $\mathcal{I}_{\rm interf}(i)$ denote the set of terminals that interfere the transmission of child $i$
 to its parent using the test in~\eqref{eq:SINR_one}. For exposition purposes,  set $\mathcal{I}_{\rm interf}(i)$
   also includes  child  $i$ itself,  $i$'s  parent, and excludes sink terminal: 
\begin{align}
\mathcal{I}_{\rm interf}(i)  \triangleq  & ~ \bigg \{  \bigg\{ i' \in \mathcal{N}_{\rm pint}(j)    : 
  {\rm SINR}_{i \rightarrow {\rm par}(i)}^{ i'} < \theta \bigg \}  \nonumber \\
 & \cup  i \cup {\rm par}(i) \bigg \} \big \backslash \mathtt{s}.
\label{eq:Interf_Set}
\end{align}
For example, consider the network of Fig.~\ref{fig:simple_net} and a threshold
$\theta$ such that ${\rm SINR}_{3 \rightarrow 4}^{1} < \theta $
and ${\rm SINR}_{5 \rightarrow 4}^{1} < \theta $, i.e.,  both
children $3$ and $5$ are disturbed from terminal's $1$ simultaneous transmission.
  It is noted that  $\mathcal{N}_{\rm pint}(5) = \{1\}$, while  
$\mathcal{N}_{\rm  pint}(3) = \emptyset$,  because link $(1, 3)$ exists in routing connectivity.  
According to Eq.~\eqref{eq:Interf_Set}, $\mathcal{I}_{\rm interf}(1) = \{1,3\}$,
$\mathcal{I}_{\rm interf}(2) = \{2,3\}$, $\mathcal{I}_{\rm interf}(3) = \{1,3\}$,
$\mathcal{I}_{\rm interf}(4) = \emptyset $ and $\mathcal{I}_{\rm interf}(5) = \{1,5\}$.
 Also, note that WSN terminals $1$ and $5$ are connected within two hops, since $1$ and $4$ (parent of $5$)
 are in communication range, according to the interferer set detection criteria in Eq.~\eqref{eq:SINR_one} and 
WSN terminals $4$ and $5$ have a parent-child routing connection. It is emphasized again that while interference
 set detection is simplified according to the above, performance evaluation of the proposed time slot and frequency 
channel allocation algorithm will be conducted taking into account  all interferers  in Eq.~\eqref{eq:SINR_all} and not just the detected ones.

A key attribute of the proposed scheme is that specific routing
tree connectivity is assumed.
 The routing tree provides  extra structure and knowledge to  the WSN that
can be exploited by any WSN terminal. 
This (spatial) structure imposes specific child-parent connections
(Fig.~\ref{fig:simple_net}) that impose further design constraints:
\begin{itemize}
     \itemsep0em
     \item[1.] Siblings  cannot
     transmit to their parent at the same time slot.
     \item[2.] A child and its parent cannot transmit at the same time slot (due to the half-duplex
     constraint).
     \item[3.] A WSN terminal can tune at a single carrier frequency and transmit at a single frequency
channel (out of $K$) at a given time slot.
     \item[4.] A WSN terminal has knowledge of its up to $2$-hop neighbors. Furthermore, neighbors within
exactly $2$-hops cannot transmit at the same time slot and at the same
     frequency channel (due to the hidden terminal problem). It is
     remarked that in the latter case they could utilize
     different frequency channels.
\end{itemize}
The above constraints assume a routing tree and will be summarized
as \emph{routing connectivity} constraints. 

Interference is caused to a parent terminal receiver, when a single
or multiple terminals (that are not children of the specific parent
in the routing tree) are transmitting at the same time slot and at the
same frequency channel \emph{and} the SINR at the parent receiver,
as defined in Eq.~\eqref{eq:SINR_one}, is below a predetermined
threshold $\theta$; otherwise,  the specific
transmitter(s) will cause no interference:
\begin{itemize}
     \itemsep0em
     \item[5.] When a child transmits and there exists at least one interfering
simultaneous transmission, the interfering terminal(s) and the
child terminal must be assigned different frequency channels for
a given time slot.
\end{itemize}
The above constraint is due to {interference connectivity}.
Constraint 5. along with 2. and 3. constitute the \emph{interference connectivity} constraints.%
\footnote{It is emphasized that constraints 2. and 3. have to be encountered in both routing as
well as interference connectivity.}
When more than one interfering terminals are
discovered, then they must be allocated   to different frequency channels.

Finally, the algorithm should consider that each terminal
should transmit once during a
specific frame (of $M$ time slots, as described above):
\begin{itemize}
     \itemsep0em
     \item[6.] A WSN terminal transmits during exactly one time slot per transmission
     frame (with the exception of the sink which always receives).
\end{itemize}
The above constraint will be referenced as the \emph{transmission} constraint.

  The above criteria can be easily modified to accommodate modern wireless transmission technologies, such as those based on full-duplex
 radios or network coding, left for future work. This work solves the following problem: given the above set of constraints, as well as a given routing tree and a 
specific set of (practically discovered) interferers, offer an inference algorithm that allows all WSN terminals to
discover channel allocation (both in terms of time and frequency resources) that adheres to the constraints, while 
each WSN  terminal exchanges information with up to two-hop neighbors in communication connectivity. Such distributed algorithm is
accompanied with convergence and correctness guarantees, while computation costs are also meticulously taken into account.

\section{Distributed Joint Time/Frequency Allocation FG Algorithm}
\label{sec:alg}
\subsection{Factor Graph Modeling}
\label{sec:sub_fg_construction}

The factor graph (FG) construction
requires random variables, whereas their factor nodes  check their
dependencies and implement the constraints of the initial problem.
The dependencies between the random variables could be also offered
in terms of a matrix description, that resembles the parity check
matrix in factor graph-based coding literature (e.g., \cite{Gal:63, RiUr:08}).
Towards that goal, a set of binary variables $\left \{s_{i,m}^{(k)} \right \}$
is defined, with $i \in \mathcal{N}_{\backslash \mathtt{s}} ,\; m \in \mathcal{M}, \; k\in
\mathcal{K}$. Binary variable $s_{i,m}^{(k)}$ is called  scheduling variable
of terminal $i$ (excluding sink terminal) and denotes transmission ($s_{i,m}^{(k)}=1$) or no
transmission $(s_{i,m}^{(k)}=0)$ of transmitter $i$, at time slot
$m$ and frequency channel $k$.

Each constraint
variable is input to specific \emph{factor}  nodes that check
the validity of specific constraints and return $1$ if the constraints
are satisfied and $0$, otherwise. 
  Given that there are three kinds of
(local) constraints (i.e., routing, interference, and transmission), three
kinds of factor nodes are constructed:
\begin{itemize}
\item $\mathsf{f}$ factors (or routing factors): each $\mathsf{f}_{i,m}$ factor node is related
 to  terminal $i  \in \mathcal{N}$ at time slot
$m$ and checks the validity of routing connectivity constraints. 
The domain of each factor $\mathsf{f}$ is  given by
\begin{align}
 \mathbf{dom} \mathsf{f}_{i,m} = &\left\{  \left\{s_{i',m}^{(k')} \in \mathds{B} \right\}: i'  \in  \mathcal{N}_{\rm TwoH}(i),\;
k' \in \mathcal{K} \right\}    \nonumber
\\ =&~
 \mathds{B}^{ |\mathcal{N}_{\rm TwoH}(i)| K}.
\label{f_fac}
\end{align}

\item $\mathsf{h}$ factors (or interference factors): similarly, each $\mathsf{h}_{i,m}$ factor node
 is related to terminal  $i \in \mathcal{N}_{\backslash \mathtt{s}}$ at time slot
$m$ and checks the validity of interference connectivity
 constraints. The domain of each factor $\mathsf{h}$ is
given by
\begin{align}
 \mathbf{dom} \mathsf{h}_{i,m}  = & \left \{   \left\{s_{i',m}^{(k')} \in \mathds{B} 
\right\} : i' \in  \mathcal{I}_{\rm interf}(i), \;k' \in \mathcal{K} \right \}   \nonumber
\\ =& ~ \mathds{B}^{|\mathcal{I}_{\rm interf}(i)| K}.
\label{h_fac}
\end{align}

\item $\mathsf{t}$ factors (or transmission factors): for each terminal but sink there exists a corresponding $\mathsf{t}_i$ factor, 
$i \in \mathcal{N}_{\backslash \mathtt{s}}$, which is related
 to the validity of the transmission constraints.
The domain is given by 
\begin{equation}
\mathbf{dom} \mathsf{t}_{i}  =  \left \{   \left\{s_{i,m'}^{(k')} \in \mathds{B} \right\} :  m' \in \mathcal{M} , \;k' \in \mathcal{K}
\right \} = \mathds{B}^{K M}.
\end{equation}
\end{itemize}

Each WSN terminal (except sink) has $2M+1$ (local) factor nodes 
and $MK$ variables nodes; specifically, for each $i \in \mathcal{N}_{\backslash \mathtt{s}}$, 
there are $M$ routing connectivity factors ($\mathsf{f}_{i,m}, \; \forall m \in \mathcal{M}$), $M$
interference connectivity factors ($\mathsf{h}_{i,m},\; \forall m \in
\mathcal{M}$) and one transmission factor ($\mathsf{t}_{i}$). The sink terminal has $M$ factors and no variables.
Detailed description of $\mathsf{f,h}$ and $\mathsf{t}$ factors is offered at Appendix~\ref{Appendix:app1}.

 A few factor domain examples are given for the network of Fig.~\ref{fig:simple_net} with $M=2$ time slots and $K=2$ frequency channels:
\begin{align}
\mathbf{dom}\mathsf{ f}_{4,1}\! &~= \left\{ \mathds{B}^8 \ni  \left [ s_{1,1}^{(1)}~ s_{2,1}^{(1)}~ s_{3,1}^{(1)}~ s_{5,1}^{(1)}~s_{1,1}^{(2)}~ s_{2,1}^{(2)}~
s_{3,1}^{(2)}~ s_{5,1}^{(2)}\right ] \right\}  \nonumber , \\
\mathbf{dom}\mathsf{ h}_{5,2}\! &~= \left\{ \mathds{B}^4 \ni  \left [s_{1,2}^{(1)}~ s_{5,2}^{(1)}~
s_{1,2}^{(2)}~ s_{5,2}^{(2)}\right] \right\} , \\
\mathbf{dom} \mathsf{ t}_{3}\!&~= \left\{ \mathds{B}^4 \ni  \left[ s_{3,1}^{(1)} ~ s_{3,2}^{(1)}
~ s_{3,1}^{(2)}~s_{3,2}^{(2)}\right]  \right \} \nonumber .
\end{align}
Do note that the domain of interference connectivity factor $\mathsf{ h}_{5,2}$ includes binary variables $s_{5,2}^{(1)}$
and $s_{5,2}^{(2)}$ of WSN terminal $5$ itself.
 Also notice that the same domain of WSN terminal $5$
 includes variables from WSN terminal $1$, which is connected to $5$ in the physical WSN topology
 within two hops, due to interference connectivity, explained in the previous section \ref{sec:sys_mod}. 
In the FG bipartite topology, factor $\mathsf{ h}_{5,2}$ belongs to WSN terminal $5$ and is connected
 within 1 hop to variables $s_{1,2}^{(1)}, s_{1,2}^{(2)}$ that belong to WSN terminal $1$.
 As an additional example, consider an arbitrary input configuration
for  factor $\mathsf{f}_{4,1}$, e.g.,
 $\left[s_{1,1}^{(1)}~ s_{2,1}^{(1)}~ s_{3,1}^{(1)}~
s_{5,1}^{(1)}~s_{1,1}^{(2)}~s_{2,1}^{(2)}~ s_{3,1}^{(2)}~
s_{5,1}^{(2)}\right]=[1~ 0~ 0~0~ 0~ 0~0~ 1]$, which simply states
that terminals $1$ and $5$ transmit simultaneously at time slot $1$,
at different frequency channels; hence $\mathsf{f}_{4,1}([1~ 0~ 0~ 0~ 0~ 0~
0~1]) = 1$, since no routing connectivity  constraint is violated
from the perspective of terminal $4$ at time slot $1$.
Similarly, consider the local factor $ \mathsf{h}_{5,2}$, with configuration
$[s_{1,2}^{(1)}~ s_{5,2}^{(1)}~
s_{1,2}^{(2)}~ s_{5,2}^{(2)}]$ $= [1~ 0~0~ 1]$,
which indicates that terminals $1$ and $5$ both transmit  at time
slot $2$, at different frequency channels; thus, the reception of
$4$ is  not  interfered, hence  $\mathsf{h}_{5,2}([1~ 0~ 0~ 1]) = 1$.

\begin{figure}[t]
    \centering
    \includegraphics[width=0.6\columnwidth]{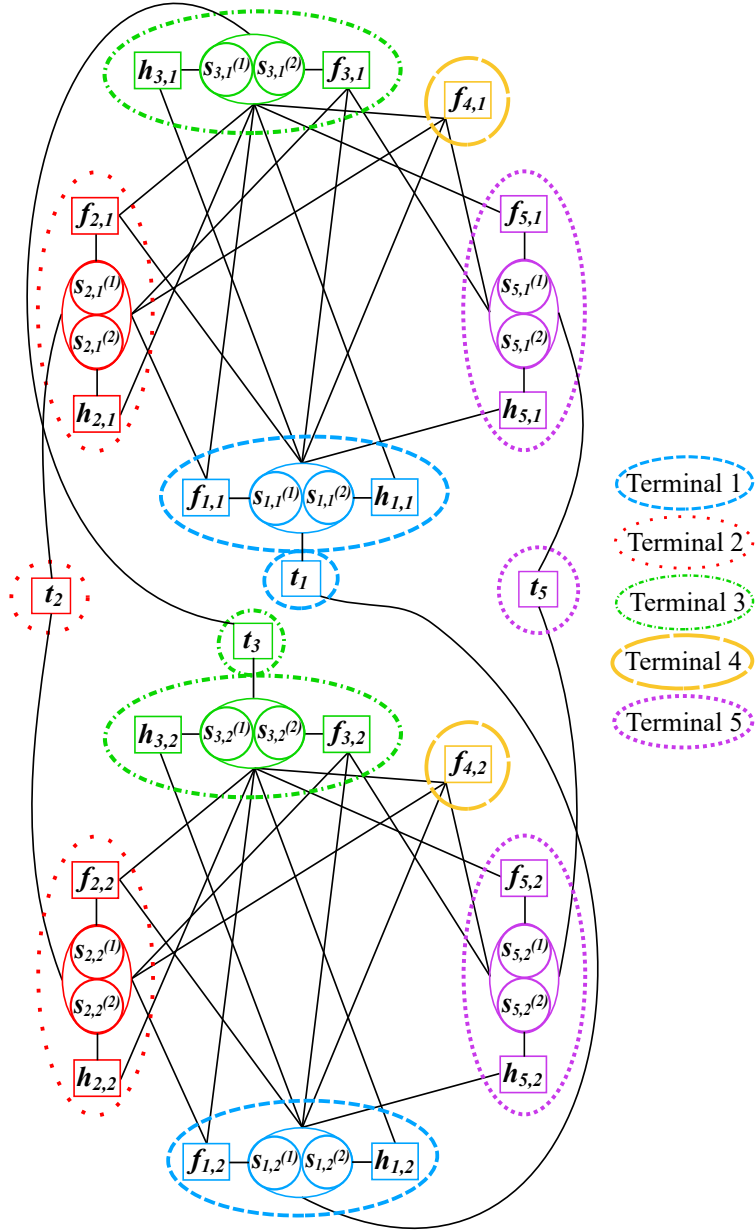}
    \caption{The factor graph (FG) corresponding to the WSN of Fig.~\ref{fig:simple_net} for $M = 2$ time slots and $K = 2$ frequency channels. For presentation purposes each depicted hyper-variable incorporates two variables and each depicted hyper-edge incorporates two edges.}
    \label{fig:FG}
\end{figure}

For each WSN terminal, the scheduling variables are constructed and
connected to the local routing connectivity, interference
connectivity, and transmission constraint factors.\footnote{The terms variable node and variable, as well as
factor node and factor will be considered equivalent subsequently.}
The goal is to find a proper time slot and frequency channel allocation that adheres to all constraints; that is equivalent to construct a FG with factorization that satisfies
\begin{align}
& \left( \prod_{i \in \mathcal{N}_{\backslash \mathtt{s}}} \mathsf{t}_{i}(\mathbf{s}_{\mathsf{t}_{i}})\right)  \cdot \left(
 \prod_{m \in \mathcal{M}} \left(  \prod_{i \in \mathcal{N}_{\backslash \mathtt{s}}}    \mathsf{h}_{i,m}\!\!\left(\mathbf{s}_{\mathsf{h}_{i,m}} 
\right)   \right)   \cdot\right. \nonumber \\
& ~~~~~~~~~~~~~~~~~~~~~~~~~~~~~~~\cdot \left.
 \left(  \prod_{i \in \mathcal{N}}
\mathsf{f}_{i,m}\!\!\left(\mathbf{s}_{\mathsf{f}_{i,m}}\right) \right)    \right)  =  1,
\label{FG_eq}
\end{align}
where $\mathbf{s}_{\mathsf{t}_{i}}, \mathbf{s}_{\mathsf{f}_{i,m}}$, and $\mathbf{s}_{\mathsf{h}_{i,m}}$ denote the variable subsets of
the corresponding factors. The appropriate value of $M$ depends on the overall (routing and interference) network connectivity, as well as the amount of traffic. In this work, $M$ is assumed fixed and chosen as the maximum node degree of the routing tree.\footnote{
At the simplest case, one terminal receives from all its children and then forwards to its parent.} When the specific choice of $M$ does not offer a valid solution, it is increased by one until a valid solution is found.
The FG of Fig.~\ref{fig:FG} corresponds to the network of Fig.~\ref{fig:simple_net} for $M=2$, $K=2$ and interference connectivity according to
 ${\rm SINR}_{3 \rightarrow 4}^{1}
< \theta$, ${\rm SINR}_{5 \rightarrow 4}^{1} < \theta $. Factor definitions offer two useful propositions that  will be exploited subsequently:

\begin{prop} \normalfont
\label{prop_f}
For  $\mathsf{f}_{i,m} $ factors,  the variable assignments
 satisfying  $\left\|\mathbf{s}_{\mathsf{f}_{i,m}}\right\|_0 \geq  |\mathcal{N}_{\rm TwoH}(i)|$
offer  $\mathsf{f}_{i,m}\!\left(\mathbf{s}_{\mathsf{f}_{i,m}}\right) = 0$.
\begin{proof}
This can be shown by translating   $ \left\|\mathbf{s}_{\mathsf{f}_{i,m}}\right\|_0 \geq  |\mathcal{N}_{\rm TwoH}(i)|$  
 in terms of  WSN routing connectivity as follows:
 if  $\left\|\mathbf{s}_{\mathsf{f}_{i,m}}\right\|_0 \geq  |\mathcal{N}_{\rm TwoH}(i)|$  then    either (a)  all $2$-hop
neighbors of $i$ in routing tree (including its child or its parent or both) transmit at the same time-slot  and the
 same frequency channel, which is inappropriate due to  constraint $2$. or   (b) 
some terminals in the   routing tree's  $2$-hop 
neighborhood  of terminal $i$ transmit at  more than $1$ frequency channels concurrently, which is inappropriate due to constraint $3$.
\end{proof}
\end{prop}

\begin{prop} \normalfont
\label{prop_h}
For $\mathsf{h}_{i,m}$ factors, the variable assignments satisfying 
 $\left\|\mathbf{s}_{\mathsf{h}_{i,m}}\right\|_0 \geq
\min\left\{ K+1, |\mathcal{I}_{\rm interf}(i) |\right\}$ offer  $\mathsf{h}_{i,m}\! \left(\mathbf{s}_{\mathsf{h}_{i,m}}\right) =  0$.
\begin{proof}
If $ |\mathcal{I}_{\rm interf}(i)|\leq K+1 \leq \left\|\mathbf{s}_{\mathsf{h}_{i,m}}\right\|_0$, then  either  (a) 
interference neighbors  (including node $i$)  transmit concurrently
with $i$'s parent, which is inappropriate according to constraint 2.   and constraint 5.  or   (b) 
  some of terminal's $i$ interferers  
 transmit at  more than $1$ frequency channels concurrently, which is inappropriate according to constraint 3.
 If $ \left\|\mathbf{s}_{\mathsf{h}_{i,m}}\right\|_0 \geq  |\mathcal{I}_{\rm interf}(i)| >  K+1$, at most $K$ frequency channels have been assigned
to  at least  $K+2$ terminals; that is inappropriate due to  constraint 3.  
\end{proof}
\end{prop}

\subsection{Proposed Synchronous Loopy BP}

\label{subsec:proposed_loopy_BP}

For exposition purposes of the loopy BP algorithm, simplified notation for variable and factor nodes is adopted. Specifically,  the set of variables is relabeled as:
\begin{equation}
\left \{ s_{i,m}^{(k)}:  i \in \mathcal{N}_{\backslash \mathtt{s}},
  m \in \mathcal{M}, k \in \mathcal{K} \right\} \! \triangleq\!
\{x_{1},x_{2},\ldots, x_{M  K (N-1)}\}
\end{equation}
and each variable is indexed by elements in the following set: 
\begin{align}
\mathcal{V} \triangleq \{ 1 , 2 ,\ldots, M  K (N-1)  \}.
\label{eq:variable_index_set}
\end{align} 
In that way, for every $v \in \mathcal{V}$ there exist unique $i \in \mathcal{N}_{\backslash \mathtt{s}}$,
 $m \in \mathcal{M}$ and  $k \in \mathcal{K}$, 
such that $x_v = s_{i,m}^{(k)}$.

Moreover, the set of factors is relabeled as follows:
\begin{align}
&\left\{ \left\{\mathsf{f}_{i,m} \right\}_{(i,m) \in \mathcal{N}\times\mathcal{M} }
, \left\{\mathsf{h}_{i,m} \right\}_{(i,m) \in \mathcal{N}_{\backslash \mathtt{s}}\times\mathcal{M} } , 
     \left\{\mathsf{t}_{i} \right\}_{i \in \mathcal{N}_{\backslash \mathtt{s}} } \right\} \nonumber \\
\triangleq   &~ 
\{\mathsf{g}_{J}\}_{J=1}^{  (N-1)(2M+1)+M }.
\end{align}
We index the set of factors   as follows:
\begin{equation}
\mathcal{J} \triangleq \{  1, 2,\ldots, (N-1)(2M+1)+M \}.
\label{eq:factor_index_set}
\end{equation} 
As a result, for any factor  $\mathsf{g}_J$, $J \in \mathcal{J}$, there exist unique $i \in \mathcal{N}$ 
 and $m \in \mathcal{M}$  such that $\mathsf{g}_J= \mathsf{f}_{i,m}$, or   unique $i \in \mathcal{N}_{\backslash \mathtt{s}}$ 
 and $m \in \mathcal{M}$ such that $\mathsf{g}_J= \mathsf{h}_{i,m}$, or  unique 
$i \in  \mathcal{N}_{\backslash \mathtt{s}}$ such that $\mathsf{g}_J=\mathsf{t}_i$.
A variable $x_v,v \in \mathcal{V}$, is an argument in factor $\mathsf{g}_J, J \in \mathcal{J}$,
if and only if, $x_v$ is adjacent to $\mathsf{g}_J$ in the  FG.
The neighborhood  of index variable $v \in \mathcal{V}$ and index factor $J \in \mathcal{J}$ in the FG is defined as follows:
\begin{align}
\mathcal{J}_v&\triangleq \{ J \in \mathcal{J}: \text{variable $x_v$  is adjacent to factor $\mathsf{g}_J$}\}, \\
\mathcal{V}_J&\triangleq \{ v  \in \mathcal{V}: \text{variable $x_v$  is adjacent to factor $\mathsf{g}_J$}\},
\label{eq:connectivity_factor_J}
\end{align}
respectively.  

The messages at iteration $n$ from variable nodes to
factor nodes and vice versa are denoted by $\mathsf{m}^{(n)}_{ v
\rightarrow  J}(x_v)$ and $\mathsf{m}^{(n)}_{ J \rightarrow  v}(x_v)$,
respectively. The proposed algorithm \emph{initializes independently} each message
$ \mathsf{m}^{(0)}_{ v \rightarrow J}(0) = 1 - \mathsf{m}^{(0)}_{ v \rightarrow
 J}(1) = q_{ v}$, with $q_{ v} \sim \mathcal{U}[0,1]$ and
$\mathsf{m}^{(0)}_{ J \rightarrow  v}(0) = \mathsf{m}^{(0)}_{ J \rightarrow
 v}(1) = 0$.  Parameter $q_{ v}$ can be considered as the initial
random guess (prior) of the corresponding scheduling variable being 0, i.e., $\text{P}_{ v}(0) = q_{ v} = 1 -\text{P}_{ v}(1)$
(where $\text{P}_{ v}(x), \; x\in \mathds{B}$, is the prior probability distribution function of binary variable $x_v, \; v \in \mathcal{V}$). 
 For each $n \in \mathds{N}$, the
standard  BP update rules follow \cite{KscFreLoe:98, Lol:04}:
\begin{align}
\mathsf{m}^{(n)}_{ J \rightarrow  v}(x_v) &=  \mathtt{C}_{J \rightarrow  v}^{(n)}(\mathbf{q})  \!\! \!\!\!  \sum_{\mathbf{x}_{\mathcal{V}_J \backslash{v}} 
\in \mathds{B}^{|\mathcal{V}_J |-1}} \!\! \!  \mathsf{g}_J( \mathbf{x}_{\mathcal{V}_J}  )  \!\!
\prod_{{y} \in \mathcal{V}_J  \backslash v } \!\!\mathsf{m}^{(n-1)}_{ {y} \rightarrow  J}(x_{y}),   \label{fac_up_rule_1}\\
\mathsf{m}^{(n)}_{ v \rightarrow J}(x_v ) &=  \mathtt{C}_{v \rightarrow  J}^{(n)}(\mathbf{q})~ 
\text{P}_{ v}(x_v) \prod_{I \in\mathcal{J}_v  \backslash  J} \mathsf{m}^{(n)}_{ {I} \rightarrow  v}(x_v),
\label{var_up_rule_1}
\end{align}
where constants $ \mathtt{C}_{J \rightarrow  v}^{(n)}(\mathbf{q}) $ and $\mathtt{C}_{v \rightarrow  J}^{(n)}(\mathbf{q}) $ guarantee  that 
$\mathsf{m}^{(n)}_{ J \rightarrow  v}(0) + \mathsf{m}^{(n)}_{ J \rightarrow  v}(1) = 1$ and 
$\mathsf{m}^{(n)}_{v \rightarrow  J}(0) + \mathsf{m}^{(n)}_{ v \rightarrow  J}(1) = 1$, respectively; their  value depends on a subset of priors $\mathbf{q} = [q_{ 1}~ q_{ 2}~ \ldots~ q_{ {MK(N-1)}}] 
\in [0,1]^{MK(N-1)}$,  as well as the current iteration.   
By the definition of $\mathbf{q}\subset [0,1]^{MK(N-1)}$, it can be  shown inductively that 
 $ \mathtt{C}_{J \rightarrow  v}^{(n)}(\mathbf{q}) \geq1 $ and $\mathtt{C}_{v \rightarrow  J}^{(n)}(\mathbf{q})\geq 1 $,
$\forall n \in \mathds{N}$.
  For any   variable assignment $ \mathbf{x}_{\mathcal{V}_J} \in \mathds{B}^{|\mathcal{V}_J|} =  \mathbf{dom} \mathsf{g}_J$,
 factor $ \mathsf{g}_J$ returns either one or zero. 
Notation $ \mathbf{x}_{\mathcal{V}_J  \backslash v} \in \mathds{B}^{|\mathcal{V}_J|-1} $  under the summation
indicates sum over all possible binary configurations of the variables in vector
$\mathbf{x}_{\mathcal{V}_J }$ except variable $x_v$. It is also noted that each message along an edge $(J,v)$ or $(v,J)$ can be parameterized by a single real number.

In addition, a damping technique  can be  employed  to decrease the probability of divergence  \cite{Hes:04, ShLeeVishVic:12, RanSchFl:14}.
Specifically, after the calculation of messages from factors to variables
in Eq.~\eqref{fac_up_rule_1}, the following damping step is utilized:
\begin{equation}
\mathsf{m}^{(n)}_{ J \rightarrow v}(x_v)  = \alpha^{(n)} \mathsf{m}^{(n-1)}_{ J \rightarrow  v}(x_v)\! + \!
\left(1-\alpha^{(n)} \!\right) \mathsf{m}^{(n)}_{ J \rightarrow  v}(x_v),
\label{eq:damping}
\end{equation}
with $\alpha^{(n)} \in [0, 1) ,\; \forall n \in  \mathds{N}$.
 Finally, marginals, denoted as
${\mathsf{r}}_v^{(n)}(x_v),~\forall v \in \mathcal{V}$,  determine the final values of the
scheduling random variables:
\begin{equation}
\mathsf{r}_v^{(n)}(x_v) =   \text{P}_{ v}(x_v) 
\prod_{J \in \mathcal{J}_v  } \mathsf{m}^{(n)}_{ J \rightarrow  v}(x_v) .
\label{marginal}
\end{equation}
The value of each variable at iteration $n$ is inferred using the following rule:
\begin{equation}
\widehat{x}^{(n)}_v  = \mathsf{1}\!\left\{ \mathsf{r}_v^{(n)}(1) \geq  \mathsf{r}_v^{(n)}(0) \right\},
~~\forall v \in \mathcal{V}.
\label{eq:MAP_rule}
\end{equation}
 Denote
$\widehat{\mathbf{x}}^{(n)} \triangleq \left\{ \widehat{x}^{(n)}_v\right\}_{v \in \mathcal{V}}$
and define the following:
\begin{equation}
\mathsf{FG}\!\left(\widehat{\mathbf{x}}^{(n)}\right)
\triangleq \prod_{J \in \mathcal{J}} \mathsf{g}_J\!\left( \widehat{\mathbf{x}}^{(n)}_{\mathcal{V}_J}\right).
\end{equation}
In a centralized implementation the algorithm terminates at the first iteration index $n^{\star}$,
for which    $\mathsf{FG}\!\left(\widehat{\mathbf{x}}^{(n^{\star})}\right)=1$.
In a distributed implementation,  the algorithm    terminates
after  a predetermined number $N_\text{iter}$ of iterations.

Finally, it is emphasized that the calculation  
of each outgoing message (across an  FG edge) at WSN terminal $i$ requires
reception of incoming messages from WSN terminals $j \in  \mathcal{N}_{\rm TwoH}(i) \cup \mathcal{I}_{\rm interf}(i)$.

\subsection{Extension to Asynchronous Scheduling}
\label{subsec:async_scheduling}

The update rules in Eqs.~\eqref{fac_up_rule_1}
and~\eqref{var_up_rule_1} can be  modified  to adhere to an
 asynchronous scheduling \cite[Chapter 6]{BertTsi:1989}.
Specifically,  
let $\{(t_n)_n; n \in \mathds{N} \cup \{0\} \}$ be the time instants at which the 
outgoing message across an arbitrary  edge $(J,v)$ is calculated. The sequence
is increasing,  goes to infinity, and $t_0 = 0$.
The outgoing message at the $n$th step, $\mathsf{m}^{(t_n)}_{ J \rightarrow  v}(x_v)$, 
 can be computed using  the most up-to-dated values of 
incoming messages. 
Let  $\{t'_{{y} \rightarrow  J}(t_n)\}_{y \in \mathcal{V}_J  \backslash v}$
  be the  time indexes  of the most up-to-dated  values of incoming messages,
and all of them are smaller than $t_n$.
Then, under an asynchronous scheduling, the update rule in  Eq.~\eqref{fac_up_rule_1}
for the $n$th step can be computed using messages values
 $\left\{\mathsf{m}_{{y} \rightarrow  J}^{(t'_{{y} \rightarrow  J}(t_n))} (x_{y}) \right\}_{ {y} \rightarrow  J}$.
Similar reasoning can be applied to the calculation of variable-to-factor update rules, as well as  to
the damped version of BP.

\section{Convergence and Complexity}
\label{sec:convergence_complexity}

\subsection{Convergence Sufficient Condition}
\label{sec:convergence}

This section offers sufficient conditions for convergence to a valid solution, despite the loopy nature of the crafted FG.
 The offered theorem also assisted in modifications of the MP procedure that accelerate convergence, discussed below.

 
The following quantity is defined for all   $v \in \mathcal{V}$:
\begin{equation}
\kappa_v \triangleq   \max_{J \in \mathcal{J}_v} \left| \left\{  
\mathbf{x}_{\mathcal{V}_J} \!  \in\! \mathds{B}^{|\mathcal{V}_J|} \! :\mathsf{g}_J(\mathbf{x}_{\mathcal{V}_J})\!=\!1
\right\} \right|.  \label{eq:k1}
\end{equation}
It is noted that $|\mathcal{J}_v| \geq 3$, due to the fact that each variable $x_v$ (associated with a variable $s_{i,m}^{(k)}$, $i \in \mathcal{N}_{\backslash \mathtt{s}}$) has
at least three adjacent factor nodes in the crafted FG: factor $\mathsf{f}_{i,m}$, factor $\mathsf{h}_{i,m}$, and factor  $\mathsf{t}_i$.
Additionally, $\mathsf{\kappa}_v \geq 1$ holds, due to the definition of factor nodes in~Appendix~\ref{Appendix:app1}: there exists at least one configuration in their domain offering $\mathsf{g}_J(\cdot) = 1$.

The following theorem exploits the structure of the crafted FG and shows that if a valid solution exists, appropriate initialization of priors $\{q_{v} \equiv \text{P}_{ v}(0)=1-\text{P}_{ v}(1)\}_{v \in \mathcal{V}}$ guarantees convergence (to that solution) of the loopy BP algorithm:

\begin{theor} \normalfont
\label{theorem:convergence_valid_solution_every_n_natural}
Suppose that there is at least one valid solution $ \boldsymbol{x}^{\star}  =\left [\mathtt{x}_1^{\star}~ \mathtt{x}_2^{\star}~ \ldots~
  \mathtt{x}_{MK(N-1)}^{\star} \right]$. Constants $\{\epsilon_v\}_{v \in \mathcal{V}}$ are defined, so that they solely depend on the
crafted FG, which in turn is associated with the WSN topology:
\begin{equation}
\epsilon_v \triangleq \frac{1}{1 + (\kappa_v)^{\left|  \mathcal{J}_v \right|}}\in \left(0,\frac{1}{2}\right] .
\label{epsilon_val}
\end{equation}
Sufficient condition for the loopy BP  algorithm to offer solution
$\widehat{x}^{(n)}_v = \mathtt{x}_v^{\star}  , ~ \forall v \in \mathcal{V},\; \forall n\in \mathds{N}$,
is the following initialization of the priors $\{q_{v} \equiv \text{P}_{ v}(0)=1-\text{P}_{ v}(1)\}_{v \in \mathcal{V}}$:
\begin{equation}
 \begin{array}{ll}
 1\geq q_{ v}   > 1 - \epsilon_v, ~~  \text{if}~   \mathtt{x}_v^{\star} = 0 \cr  
 0 \leq q_{ v}   < \epsilon_v,  ~~~~~~~ \text{if} ~   \mathtt{x}_v^{\star} = 1.  
\end{array}
\label{priors_vals_sufficiency}
\end{equation}
In other words, if vector   $\mathbf{q} = \{q_{v} \}_{v \in \mathcal{V}}$ satisfies Eq.~\eqref{priors_vals_sufficiency}, then 
loopy BP algorithm offers exact solution  $\boldsymbol{x}^{\star}, \forall n \in \mathds{N}$, i.e.,
\begin{equation}
\mathsf{r}_v^{(n)}\!\left( \mathtt{x}_v^{\star} \right)   >  \mathsf{r}_v^{(n)}\!\left(1- \mathtt{x}_v^{\star}\right),\;
 \forall v\in \mathcal{V},\;\forall n \in \mathds{N}.
\label{eq:marginal_final_forall_n}
\end{equation}
\end{theor}
\begin{IEEEproof}
See Appendix~\ref{Appendix:app3}.
\end{IEEEproof}

\subsection{Convergence Acceleration}
\label{sec:convergence_accel}

Theorem~\ref{theorem:convergence_valid_solution_every_n_natural} states that there are prior values for $\{q_{v}\}_{v \in \mathcal{V}}$ that guarantee convergence to a valid solution, when such solution exists. 
The fact that every variable node $x_v$ in the FG is aware of its prior value $q_{v}$, 
 motivates us to perform a slight modification of the sum-product/BP procedure of Section~\ref{subsec:proposed_loopy_BP}.
A \emph{periodic} check of the problem constraints is conducted, i.e., the value of each FG factor node is tested locally every $N_\text{interm}$ iterations, using as input
the estimated values of its connected variable nodes. 
If   output value is $0$ (corresponding constraint is \emph{not} satisfied) then a \emph{flag} message
 is transmitted to the neighboring (to that factor) variable nodes. In that case, all such variables nodes re-initialize
 their priors randomly and  the iterative calculations associated with
that factor will be restarted. In short,  for any $l \in \mathds{N}$ such that 
$l \, N_\text{interm} \leq N_\text{iter}$, each local factor $\mathsf{g}_{J},\,J \in \mathcal{J}$ sends a flag  message to neighboring variables $x_v, v\in \mathcal{V}_J$ if:
\begin{equation}
\mathsf{g}_{J} \!\left(\widehat{\mathbf{x}}^{(l \, N_\text{interm})}_{\mathcal{V}_J}\right ) = 0.
\end{equation}
In that case, these variables re-initialize their  priors $q_v \sim \mathcal{U}[0,1]$,  \mbox{$ v\in \mathcal{V}_J$}. 
It is emphasized that such flag message above involves only neighboring radio terminals,
due to the specific problem formulation (and the corresponding FG formation). Numerical
results showed that the above modification accelerated convergence to a valid solution.

\subsection{Complexity Tradeoff and Computational Cost Reduction}
\label{subsec:comp_compl}
The computation cost of sum-product in Eq.~\eqref{fac_up_rule_1} is exponential with the factor node degree. For example, in the FG of Fig.~\ref{fig:FG}, the computational cost per iteration is
dominated by the update rules of factors $\mathsf{f}_{3,m},\mathsf{f}_{4,m},  m=1,2$, each with degree $8$,
 requiring  calculating operations in the order of $2^{8}$, for each
factor-outgoing message. The degree of each factor node is solely
determined by: a) the density of routing/interference links (where density of interference links
 depends on $\theta$) and b) the number of available orthogonal frequency channels (as can be seen from Eqs.~\eqref{f_fac}
and~\eqref{h_fac}).

By choosing large $\theta$, the
receivers require higher SINR and interference
connectivity is enriched; thus, algorithmic complexity is also increased. In that case, the
algorithm operates under stringent constraints and if a solution is
found, it will offer lower remaining interference compared to the case of smaller $\theta$. However,
computational time is increased. On the contrary, smaller $\theta$
reduces the number of interfering terminals and hence, the offered
solution will provide higher remaining interference and thus, weaker
overall network performance. However, computational complexity and required
time is decreased. Therefore, the overall algorithm offers an
interesting performance/complexity tradeoff, through the choice of $\theta$.

\subsubsection{Algorithmic Developments for Reduced Computational Cost}
\label{subsubsec:improvements_cost}

In order to reduce computations associated with  factor $\mathsf{g}_J$, $J \in \mathcal{J}$ in the summation 
of Eq.~\eqref{fac_up_rule_1}, which in principle involves $2^{|\mathcal{V}_J|}$ variable configurations for each iteration
$n$, the following sets are defined, for each FG edge $(J,v)$, $v \in \mathcal{V}_J$:
\begin{align}
\!{\mathcal{X}}_{(J,v)}^{1} (x) \!&\triangleq \! \left \{\mathbf{x}_{\mathcal{V}_J \backslash v}
\! \in \! \mathds{B}^{|\mathcal{V}_J|-1}\! \! : \!\mathsf{g}_J( \mathbf{x}_{\mathcal{V}_J \backslash v}, x_v =x) \!=\! 1 \right \}\! ,
\label{X_1_minus_delta} \\
\!{\mathcal{X}}_{ (J,v)}^{ 0} (x)\!&\triangleq \!  \left \{\mathbf{x}_{\mathcal{V}_J \backslash v}
\! \in\! \mathds{B}^{|\mathcal{V}_J|-1} \!  \!: \! \mathsf{g}_J( \mathbf{x}_{\mathcal{V}_J \backslash v}, x_v = x)\! =\! 0 \right \} \!.
\label{X_delta}  
\end{align}
For any edge $(J,v)$,
the set $\mathcal{X}_{(J,v)}^{0}(x)$   does not contribute in the summation
of sum-product in Eq.~\eqref{fac_up_rule_1}.
Hence,   it suffices
at each iteration $n$ to evaluate Eq.~\eqref{fac_up_rule_1}  by summing  
all $\left\{ \prod_{{y} \in \mathcal{V}_J  \backslash v } \!\mathsf{m}^{(n-1)}_{ {y} \rightarrow  J}(x_{y})\right \}$
  for only the assignments in set of Eq.~\eqref{X_1_minus_delta}.
A  binary-tree search \cite{RusNor:10}  can be further utilized in order
to pre-compute efficiently the set  in~\eqref{X_1_minus_delta}
and avoid exhaustive enumeration. 
The following proposition shows that the set of valid
assignments can be significantly smaller subset of the FG factor nodes' domain.
\begin{prop} \normalfont
\label{prop:cardinality_of_assignements}
For a multi-channel scenario with $K\geq 2$ and $\mathsf{g}_J = \mathsf{f}_{i,m}$  
  for some $i\in \mathcal{N}$ and $m \in \mathcal{M}$,
the number of valid assignments offering $\mathsf{g}_J(\cdot)=1$, i.e., set cardinality $\left|{\mathcal{X}}_{(J,v)}^{1}(0) \cup {\mathcal{X}}_{(J,v)}^{1}(1)\right|$ for some $v \in \mathcal{V}_J$,
is upper bounded by $V_{\mathsf{f}_{i,m}} =  2^{\mathsf{H}(\delta_{\mathsf{f}_{i,m}})K|\mathcal{N}_{\rm TwoH}(i)|}$,
with $\delta_{\mathsf{f}_{i,m}} = \frac{|\mathcal{N}_{\rm TwoH}(i)|-1}{K|\mathcal{N}_{\rm TwoH}(i)|}$ and
   $\mathsf{H}(x) \triangleq -x \mathsf{log}_2(x) - (1-x)\mathsf{log}_2(1-x)$.
Similar result can be obtained for $\mathsf{g}_J = \mathsf{h}_{i,m}$, where
$V_{\mathsf{h}_{i,m}} = 2^{\mathsf{H}(\delta_{\mathsf{h}_{i,m}})K|\mathcal{I}_{\rm interf}(i)|}$,
with $\delta_{\mathsf{h}_{i,m}} = \frac{ {\rm min}\{ K, |\mathcal{I}_{\rm interf}(i)|-1\}}{K|\mathcal{I}_{\rm interf}(i)|} $.
\begin{proof}
 According to Proposition \ref{prop_f}, the assignments  $\mathbf{s}_{\mathsf{f}_{i,m}} \in \mathbf{dom} \mathsf{f}_{i,m}
 = \mathds{B}^{|\mathcal{N}_{\rm TwoH}(i)| K}$  satisfying 
 $\left\|\mathbf{s}_{\mathsf{f}_{i,m}}\right\|_0  \geq |  \mathcal{N}_{\rm TwoH}(i)  |$ 
 offer $\mathsf{f}_{i,m}(\mathbf{s}_{\mathsf{f}_{i,m}})= 0$.
 Thereby, the assignments $\mathbf{s}_{\mathsf{f}_{i,m}} \in \mathbf{dom} \mathsf{f}_{i,m}$
  that offer  $\mathsf{f}_{i,m}(\mathbf{s}_{\mathsf{f}_{i,m}})= 1$
    have   $\left\|\mathbf{s}_{\mathsf{f}_{i,m}}\right\|_0$ strictly  less than  $|\mathcal{N}_{\rm TwoH}(i)|$.
 This shows that the number of valid assignments $\mathbf{s}_{\mathsf{f}_{i,m}} \in \mathbf{dom} \mathsf{f}_{i,m}$,
  cannot be more than  $\sum_{j=0}^{|\mathcal{N}_{\rm TwoH}(i)|-1}
{|\mathcal{N}_{\rm TwoH}(i)| K \choose j}$. Using the result in \cite[Lemma~16.19]{FlGr:06}
to upper bound the sum of binomial coefficients,
we obtain $\sum_{j=0}^{|\mathcal{N}_{\rm TwoH}(i)|-1}
{|\mathcal{N}_{\rm TwoH}(i)| K \choose j} 
\leq   V_{\mathsf{f}_{i,m}}$.
Exactly same reasoning can be followed for $\mathsf{g}_J = \mathsf{h}_{i,m}$ with the help of Proposition \ref{prop_h}. 
 \end{proof}
\end{prop}
It is noted that for large $K$, $V_{\mathsf{f}_{i,m}}  \ll |\mathbf{dom}  \mathsf{f}_{i,m}| ={2}^{|\mathcal{N}_{\rm TwoH}(i)|K} $
and $V_{\mathsf{h}_{i,m}}  \ll |\mathbf{dom}  \mathsf{h}_{i,m}| ={2}^{|\mathcal{I}_{\rm interf}(i)|K} $. Thus,
 for   edge $(J,v)$, only a small subset of 
 assignments contribute in the summation of Eq.~\eqref{fac_up_rule_1} and can be pre-computed and stored efficiently.

\section{Numerical Results}
\label{sec:results}

\begin{figure}[!t]
      \centering
     \includegraphics[width=0.6\columnwidth]{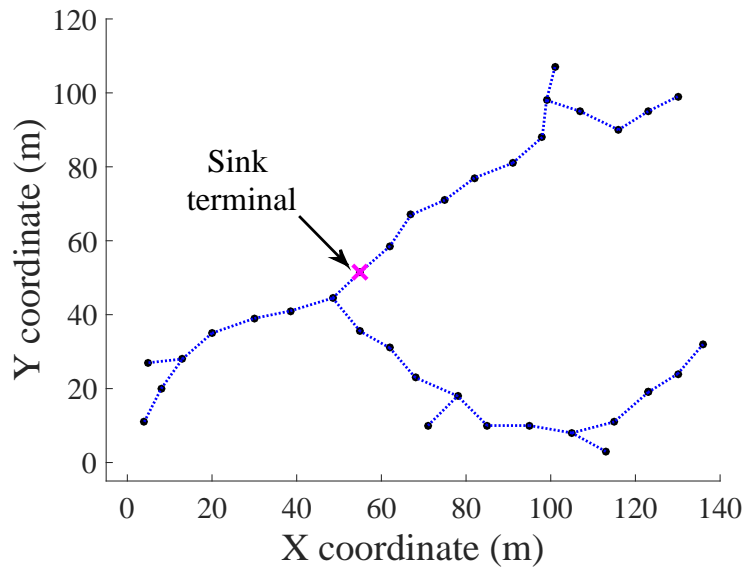}
    \caption{Routing connectivity of a multi-hop, $34$-terminal   WSN. 
Sink terminal is depicted with 'x' marker.
Overall connectivity among WSN terminals depends on radio sensitivity and propagation environment and also includes interference connectivity. }
           \label{fig:big_topology}
\end{figure}

The distributed frequency allocation algorithms GBCA \cite{CheYuChFaSh:11} and MinMax
\cite{SaiXuLuChen:11}, as well as the proposed FG-based frequency allocation algorithm (FG),
have been simulated in the Castalia network simulator \cite{Bou:0};
the latter is based on the OMNET++ platform \cite{lnk1:0}. The
Tunable MAC module of Castalia has been modified as described in
\cite{Vla:12}.
The lognormal shadowing model is adopted for radio propagation 
\cite{Lognormal:04}, with parameter values given in
Table~\ref{ChannelValues}. The 34-terminal topology with WSN
routing connectivity in Fig.~\ref{fig:big_topology} is tested. It is assumed that
the $7$ leaf terminals generate packets with constant
bit-rate.

\begin{table}[!t]
\centering \caption{Simulation parameter values in Castalia}
\subtable[]{
\begin{tabular}{|l|c|}
  \hline
   Path-loss exponent    & $2.4$ \\
   Shadowing variance & $0$ dB\\
    Ref. distance    & $1$ m \\
   Path-loss & $55$ dB \\
  \hline
\end{tabular}
\label{ChannelValues}
}
\subtable[]{
\begin{tabular}{|l|c|}
  \hline
    $M$     & $4$ slots \\
  Time slot duration                & $10$ ms \\
  Constant bitrate     & $5$ pps \\
  Total packet size         & $312$ bytes \\
  Tx\_power         & $-10$ dBm \\
  Simulation time       & $600$ sec \\
  \hline
\end{tabular}
\label{OtherValues}
}
\end{table}

For the BP algorithm,   the maximum number of BP iterations was set to $N_\text{iter} = 50$,
 and the checking period was set to $N_\text{interm} = 8$ and $\alpha^{(n)} = 0.3,\; \forall n \in  \mathds{N}$.
As discussed in Section~\ref{subsubsec:improvements_cost}, our implementation
utilizes  the binary search  for the pre-computation of
 set ${\mathcal{X}}_{(J,v)}^{1}(x) $ in~\eqref{X_1_minus_delta},
$\forall   J \in \mathcal{J},\;
\forall v \in \mathcal{V}_J$, $x \in \mathds{B}$.
For the  WSN topology of Fig.~\ref{fig:big_topology}, classic BP ($N_\text{interm} = \infty$)
did never converge within   $N_\text{iter} = 50$ iterations.

 The $34$-terminal  multi-hop network of Fig.~\ref{fig:big_topology} has relative sparse interference connectivity,
 since  every terminal can hear from $3$ to $8$ transmissions.
 Routing links offer received SNR values that exceed $-90$ dBm
and no retransmissions are allowed.
Other parameters are given in
Table~\ref{OtherValues} and simulation implementation details for
the GBCA and MinMax algorithms can be found in \cite{Vla:12}.

\begin{figure}[!t]
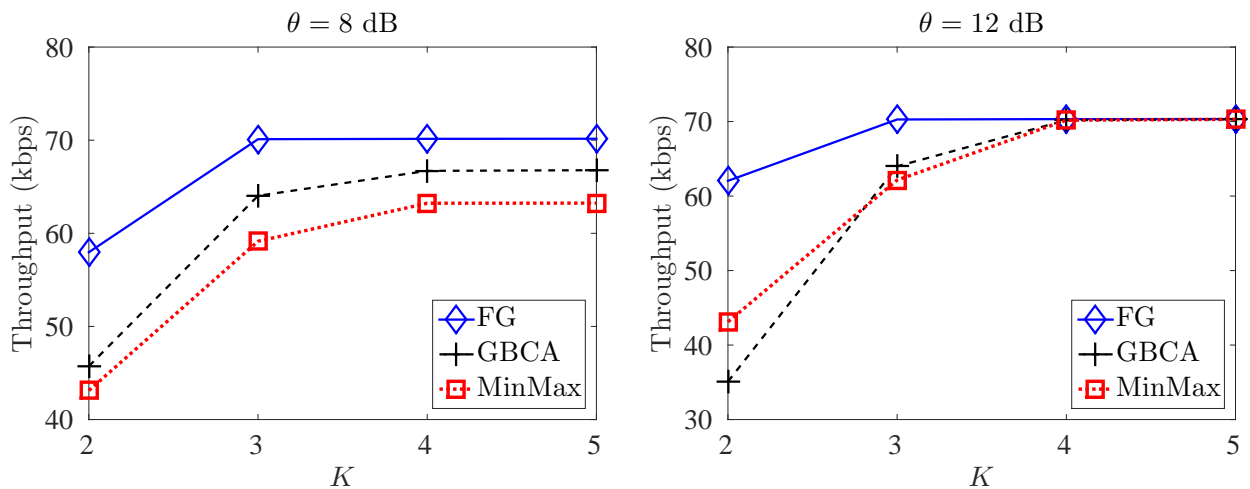

       \centering
      \hspace{-0.12 in}
            \includegraphics[width=0.48\columnwidth]{throughput_theta_8_v4} \quad
              \includegraphics[width=0.48\columnwidth]{throughput_theta_12_v4}
    \caption{Left (Right): Throughput versus number of frequency  channels, for   SINR receiver threshold 
    $\theta = 8$dB ($\theta = 12$dB) and network of Fig.~\ref{fig:big_topology}.}
     \label{fig:throughput_topology}
\end{figure}

Fig.~ \ref{fig:throughput_topology} illustrates the throughput  as a function of available frequency
channels ($K$) and receiver SINR threshold $\theta$, for the  topology in Fig.~\ref{fig:big_topology}. 
The latter is receiver-dependent and controls  the number of detected interfering  links;
 higher $\theta$ results to larger number of detected interfering links, better interference reduction and higher computational cost. It can be seen that as the number of available frequency
channels increases, higher throughput is achieved
for all protocols as expected, since frequency channel availability
reduces or eliminates interference. Fig.~\ref{fig:throughput_topology} shows that for all $\theta $
no algorithm achieves $100 \%$ packet delivery ratio, or equivalently $87.36$ kbps throughput performance.%
\footnote{$7$ flows $\times  5$ pps $\times 312$
 bytes $\times 8$ bits/byte$=87.36$ kbps.} This stems from the fact that for the  specific topology, 
 received SNR in routing tree links was relatively small due to the large distances, compromising reliability
 and packet delivery ratio.

Interestingly, for $\theta=8$ dB,    FG method outperforms
the other two algorithms  in terms of throughput, for any number  of available channels.
From Fig.~\ref{fig:big_topology} we note that for   $K\geq3$,  the throughput
of the FG method is not improved significantly and stays almost fixed. 
It can be  seen that for $K=2$, FG offers a throughput gain of  $27\%$ 
compared to GBCA and   $34\%$  compared to MinMax.
In Fig~\ref{fig:throughput_topology}-Right
the throughput performance for  $\theta=12$ dB is depicted.
It is noted that the throughput performance of all algorithms  becomes the same
for $K \geq 4$ in both cases.
For $\theta=12$ dB, the maximum throughput gain of FG over GBCA  and MinMax is      $76\%$  
and  $44\%$, respectively. 
 In  all examined cases, the maximum  packet delivery ratio was approximately $80 \%$.
The superiority of FG method  stems from the fact that
frequency and time allocation is jointly applied during the algorithm,
offering more degrees of freedom to eliminate the interference.
In contrast,  the other two algorithms divide the  time scheduling and frequency 
assignment in separate phases during their execution.

\begin{figure}[!t]
      \centering
    \includegraphics[width=0.6\columnwidth]{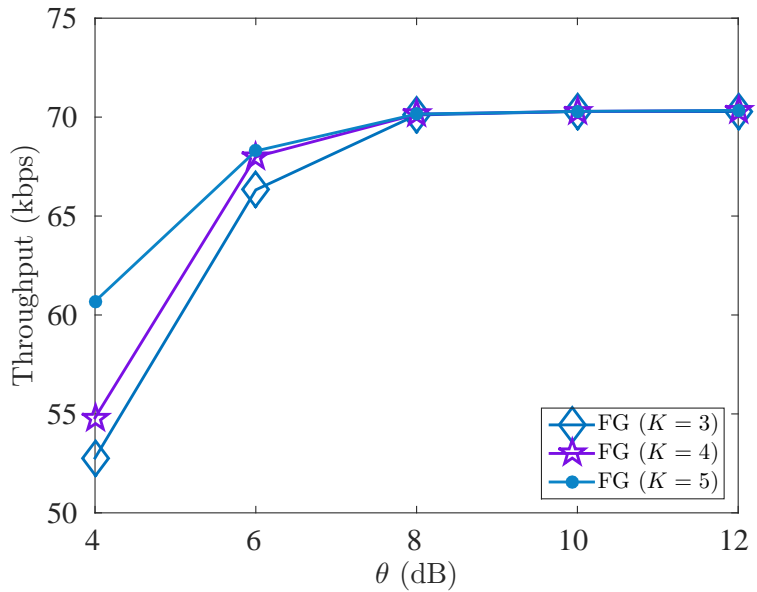}
    \caption{FG throughput versus receiver SINR threshold $\theta$ for different  number of frequency channels for network of  Fig.~\ref{fig:big_topology}.}
    \label{fig:throughput_vs_theta}
\end{figure}

Fig.~\ref{fig:throughput_vs_theta} examines the impact of number of
available frequency channels $K$ on throughput performance over different values of threshold $\theta$.
It is noted that as  the value of SINR parameter $\theta$ increases, throughput is also
increased. The proposed algorithm outperforms  both GBCA and MinMax in all cases.
We include for clarity only the proposed FG algorithm and observe that parameter threshold $\theta = 8$ dB is sufficient for the FG
algorithm to eliminate interference, offering total throughput of
approximately $70$ kbps.  This is highly encouraging given that the proposed FG methodology exploited a simplified interference set detection 
only among terminals with transmissions that can be heard and decoded, as opposed to several smaller received power transmissions which 
cannot be properly received, but their aggregate sum may be non-negligible. Nevertheless, it is again emphasized that simulations were
performed adhering to the natural physics of interference, where whichever WSN terminal transmitted within the same frequency channel
and time slot was taken into account in SINR and respective network performance evaluation. Thus, the FG can in principle reduce but not eliminate remaining interference. 

\begin{figure}[!t]
      \centering
       \includegraphics[width=0.6\columnwidth]{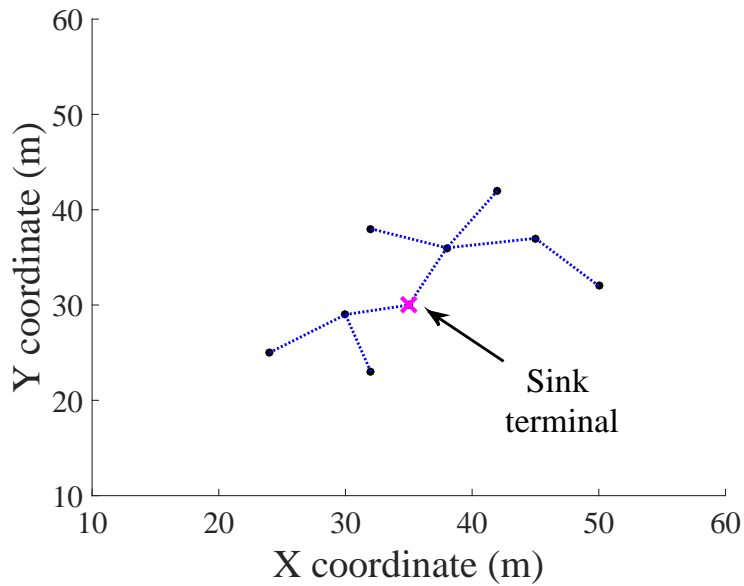}
    \caption{A 3-hop, 9-terminal  WSN topology utilized for convergence failure (outage) evaluation. }
           \label{fig:small_topology}
\end{figure}

\begin{figure}[!t]
      \centering
    \includegraphics[width=0.61\columnwidth]{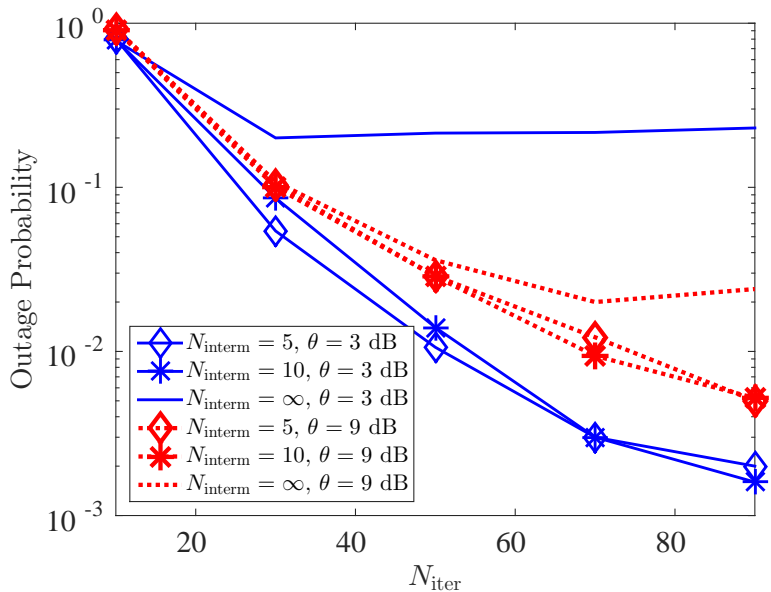}
    \caption{Outage probability of convergence to a valid solution VS maximum number of iterations,
$N_{\rm iter}$, for the 9-terminal, 3-hop topology of Fig.~\ref{fig:small_topology} and various values of  $N_{\rm interm}$. }
    \label{fig:outage_prob}
\end{figure}

Finally,  the 9-terminal, 3-hop topology of Fig.~\ref{fig:small_topology} is considered.
In Fig.~\ref{fig:outage_prob} we plot the outage probability of the proposed FG channel allocation algorithm
as a function of $N_{\rm iter}$
for $3$ different values of $N_\text{interm}$, using $K=2$ frequency channels and $\theta=3$ dB   or $\theta= 9$ dB.
The  results here are obtained by averaging over $5000$   Monte Carlo experiments.
The probability of outage is defined as follows:
\begin{equation}
\mathsf{P}_{\rm out}(N_{\rm iter}) \triangleq \text{Pr}\!\left( 
\mathsf{FG}\!\left(\widehat{\mathbf{x}}^{(N_{\rm iter})}\right)  = 0\right),
\end{equation}
i.e., the probability of FG convergence to a non-valid solution
after $N_{\rm iter}$ iterations. 
Fig.~\ref{fig:outage_prob} demonstrates that, as the number of maximum iterations
increase, the proposed modification of loopy BP decreases the probability of outage.
It is noted that the proposed modification 
of loopy BP can offer outage less than   $0.002$ for $N_{\rm iter} = 90$ with threshold $\theta=3$ dB ($0.005$ for $N_{\rm iter} = 90$
with threshold $\theta=9$ dB), in contrast to classical loopy BP,
which offers outage probability $0.2$ for  $\theta=3$ dB ($0.02$ for  $\theta=9$ dB).
Thus, we conclude that  the proposed modification in loopy BP
is in practice necessary  for probabilistic distributed channel allocation using the FG framework.
 This finding is important for network setups with high FG graph degree, either due to large $\theta$ or high WSN terminal density.

\section{Conclusion}
\label{sec:conclusion}
Factor graph-based, joint time slot/frequency channel allocation is possible in resource-constrained WSNs, 
even with truly distributed ways, i.e., local message-passing between neighboring WSN terminals,
provided that special modifications are introduced for: a) practical
 detection of interfering terminals and b) convergence of the underlying inference algorithm.
The contribution of this work is twofold: this work provides a mathematical framework 
to show convergence of loopy BP to   a valid solution, when such solution exists and the messages can 
be properly (re-)initialized; it also offers a truly distributed probabilistic algorithm that can be
implemented with realistic (even  though simplified)  practical detection of interferers. Throughput performance  
of the proposed scheme was evaluated taking into account the true nature of interference, offering promising results. 
The field of distributed resource allocation has been tremendously challenging 
and this work has barely scratched the surface, hopefully sparking further interest in the near future towards inference-based methodologies.
Modern wireless transmission technologies, e.g., based on full-duplex radio or network coding, could be
 easily incorporated by proper modifications of the constraints, left for future work.
%
%


\appendices

\section{Definition of FG Factor Nodes}
\label{Appendix:app1} %
The definition of $\mathsf{f,h}$ and $\mathsf{t}$ factors is provided below: \\
\footnotesize
\HRule \\[0.0cm]
{\textbf{function }}:~$ \mathsf{f}_{i,m}$
\label{f_func}\\ [-0.2cm]
\HRule \\[0.0cm]
{\bf Input}: $\mathbf{s}_{\mathsf{f}_{i,m}}$ \\
(1):  {\bf if} $\left(  \left\|\mathbf{s}_{\mathsf{f}_{i,m}}\right\|_0  = 0 ~~{\bf OR}~~ \left\|\mathbf{s}_{\mathsf{f}_{i,m}}\right\|_0 = 1 \right)$ \\
(2): $~~~~${\bf return } $1    ~~~$\\
(3):  {\bf else  if} $\left(  \left\|\mathbf{s}_{\mathsf{f}_{i,m}}\right\|_0  \geq |  \mathcal{N}_{\rm TwoH}(i)  |\right)$ \\
(4): $~~~~${\bf return } $0 $  // constraints 2. or 3. \\
(5):  {\bf else} \\
(6):  $~~~$ {\bf{for}}$\left\{  i_1 \in  \mathcal{N}_{\rm TwoH}(i) , k_1 \in \mathcal{K} :  {s}_{i_1,m}^{(k_1)} =  1 \right \}$  \\
(7):  $~~~$ {\bf{for}}$\left \{  i_2 \in  \mathcal{N}_{\rm TwoH}(i) , k_2 \in \mathcal{K} :  {s}_{i_2,m}^{(k_2)} =  1, [i_2,k_2]\neq [i_1,k_1] \right  \}$\\
(8): $~~~~~~$  {\bf if} $\left( k_1 \neq k_2\right)$ \\
(9): $~~~~~~~~~$  {\bf if} $\left(  i_1  \in  \left \{ \left \{ \mathcal{N}_{\rm OneH}(i_2)
 \cup   \mathcal{N}_{\rm sib}(i_2) \right\} \cap    \mathcal{N}_{\rm TwoH}(i) \right \} \right)$   \\
(10): $~~~~~~~~~~~~~~$ {\bf return } $0$  // constraints 1., 2., and  3. \\
(11): $~~~~~~~$  {\bf end if}\\
(12): $~~~~$  {\bf else}\\
(13): $~~~~~~~~$  {\bf if} $\left (i_1 \in  \left \{
\mathcal{N}_{\rm TwoH}( i_2)   \cap   \mathcal{N}_{\rm TwoH}(i)\right \} \right)$  \\
(14): $~~~~~~~~~~~~$ {\bf return } $0$ // constraint 4.  \\
(15): $~~~~~~~~$  {\bf end if}\\
(16): $~~~~$  {\bf end if}\\
(17):  $~$ {\bf{end double for}}\\
(18):  $ ~$ {\bf{return }} $1$\\
(19): {\bf end if}\\
(20): {\bf{return }} $0$\\
\HRule \\[-0.2cm]
\\ \HRule \\[0.0cm]
{\textbf{function }}:~$ \mathsf{h}_{i,m} $  
\label{h_func} \\ [-0.2cm]
\HRule \\[0.0cm]
{\bf Input}: $\mathbf{s}_{\mathsf{h}_{i,m}}$ \\
(1):  {\bf if} $\left(\left\|\mathbf{s}_{\mathsf{h}_{i,m}}\right\|_0 = 0~~ {\bf OR}~~ \left\|\mathbf{s}_{\mathsf{h}_{i,m}}\right\|_0  = 1 \right)$ \\
(2): $~~~~${\bf return } $1 $ \\
(3):  {\bf else  if} $\left(\left\|\mathbf{s}_{\mathsf{h}_{i,m}}\right\|_0 \geq {\rm min}\{ K+1, |\mathcal{I}_{\rm interf}(i)|\} \right)$ \\
(4): $~~~~${\bf return } $0 $   //  constraints 2. or 3. \\
(5):  {\bf else} \\
(6):  $~~~$ {\bf{for}}$\left\{  i_1 \in  \mathcal{I}_{\rm interf}(i) , k_1 \in \mathcal{K} :  {s}_{i_1,m}^{(k_1)} =  1 \right \}$  \\
(7):  $~~~$ {\bf{for}}$\left \{  i_2 \in  \mathcal{I}_{\rm interf}(i) , k_2 \in \mathcal{K} :  {s}_{i_2,m}^{(k_2)} =  1,
 [i_2,k_2]\neq [i_1,k_1] \right \}$\\
(8): $~~~~~$  {\bf if}  $(k_1 = k_2 )$
$~~~$  \\
(9): $~~~~~~~~$  {\bf return } $0$ // constraint 5. \\
(10): $~~~$  {\bf end if}\\
(11): $~~~$  {\bf if} $(i_1 = i_2)$\\
(12): $~~~~~~~$  {\bf return } $0$    // constraint 3.\\
(13): $~~~$  {\bf end if}\\
(14): $~~~$  {\bf if} $\left( i_1 =   i  \cap 
  i_2 = {\rm par}(i)) \right)$ \\
(16): $~~~~~~~$  {\bf return }  $0$ // constraint 2.\\
(17): $~~~$  {\bf end if}\\
(18):  $~$ {\bf{end  double for}}\\
(19): $~$  {\bf{return }} $1 $\\
(20):  {\bf{end if}}\\
(21): {\bf{return }} $0$\\
\HRule \\ [-0.2cm]
\\ \HRule \\
{\textbf{function }}:~$ \mathsf{t}_{i} $  
\label{t_func}\\ [-0.2cm]
\HRule \\[0.0cm]
{\bf Input}: $\mathbf{s}_{\mathsf{t}_{i}}$ \\
(1):  {\bf if} $\left(\left\|\mathbf{s}_{\mathsf{t}_{i}}\right\|_0  = 1 \right)$ \\
(2): $~~~~${\bf return } $1  $  \\
(3): {\bf{end if}}\\
(4): {\bf{return }} $0$\\
\HRule 
\normalsize

\section{Proof of Theorem~\ref{theorem:convergence_valid_solution_every_n_natural}}
\label{Appendix:app3} %

Two auxiliary lemmas are shown first.
\begin{lem}  \label{lemma:convergence_valid_solution_one_iteration}
\normalfont
Under the  assumptions of Theorem~\ref{theorem:convergence_valid_solution_every_n_natural},
\begin{equation}
\mathsf{r}_v^{(1)}\!\left( \mathtt{x}_{v}^{\star}\right)   >  \mathsf{r}_v^{(1)}\!\left(1- \mathtt{x}_{v}^{\star}\right),\; \;
 \forall v\in \mathcal{V},
\label{eq:marginal_final_forall_n_1}
\end{equation}
 i.e.,  the algorithm  offers outputs $\widehat{x}^{(1)}_v =  \mathtt{x}_{v}^{\star}, \,  \forall v \in \mathcal{V}$.
\end{lem}

\begin{IEEEproof}
 Suppose that prior values $\{q_v\}_{v \in \mathcal{V}}$ satisfy Eq.~\eqref{priors_vals_sufficiency}.
That initialization implies the following:
\begin{equation}
 \text{P}_{ v}(  \mathtt{x}_v^{\star})      > 1 - \epsilon_v \geq \frac{1}{2} \geq \epsilon_v >  \text{P}_{ v}( 1-  \mathtt{x}_v^{\star}),
~~~ \forall v \in \mathcal{V} .
\label{def_prior}
\end{equation}
During the first iteration, the variable nodes propagate their messages to factor nodes, in order to
calculate the outgoing messages. More specifically, $\forall v \in \mathcal{V}$ and
$\forall J \in \mathcal{J}_v$, 
\begin{subequations}\label{eq:mu_var_to_fac_iter0}
\begin{align}
\mathsf{m}^{(0)}_{ v \rightarrow  J}(0) =&~ q_{ v}=  \text{P}_{ v}(0)  \label{mu_var_to_fac_0_iter0} \\
\mathsf{m}^{(0)}_{ v\rightarrow  J}(1) =& ~1 - \mathsf{m}^{(0)}_{ v \rightarrow  J}(0) = 1- q_{ v}  = \text{P}_{ v}(1).   \label{mu_var_to_fac_1_iter0}
\end{align}
\end{subequations}

An arbitrary variable index $v_0 \in \mathcal{V}$ is chosen.  Due to the fact that factors $\mathsf{g}_J, J \in \mathcal{J}$ have  range \{0,1\}, 
 the update rule of  Eq.~\eqref{fac_up_rule_1} can be written,  for any $J \in \mathcal{J}_{v_0}$, as follows:
\begin{align}
\mathsf{m}^{(1)}_{ J \rightarrow  v_0}\!\!\left( \mathtt{x}_{v_{0}}^{\star} \right) &= \mathtt{C}_{J \rightarrow  v_0}^{(1)}(\mathbf{q}) \!\!\! \! \! \sum_{\begin{smallmatrix} \mathbf{x}_{\mathcal{V}_J} : 
 \mathsf{g}_J(\mathbf{x}_{\mathcal{V}_J}) = 1 \\ x_{v_0} = \mathtt{x}_{v_{0}}^{\star} \end{smallmatrix}} \prod_{y \in \mathcal{V}_J \backslash  v_0 }\!\! \mathsf{m}^{(0)}_{ y \rightarrow  J}(x_y),   \label{fac_to_var_iter1_bit0}\\
\mathsf{m}^{(1)}_{ J \rightarrow  v_0}\!\!\left( 1- \mathtt{x}_{v_0}^{\star} \right)  &=   \mathtt{C}_{J \rightarrow  v_0}^{(1)}(\mathbf{q}) 
\!\!\!  \! \! \sum_{\begin{smallmatrix} \mathbf{x}_{\mathcal{V}_J} :  \mathsf{g}_J(\mathbf{x}_{\mathcal{V}_J}) = 1 \\ x_{v_0}= 1 -
 \mathtt{x}_{v_0}^{\star}\end{smallmatrix}} \prod_{y \in \mathcal{V}_J \backslash  v_0 }\!\! \mathsf{m}^{(0)}_{ y \rightarrow J}(x_y).
   \label{fac_to_var_iter1_bit1}
\end{align}

 It is noted that the  summation in Eq.~\eqref{fac_to_var_iter1_bit0} scans all vectors $\mathbf{x}_{\mathcal{V}_J}$
 with $x_{v_0} = \mathtt{x}_{v_{0}}^{\star} $ that satisfy $\mathsf{g}_J(\mathbf{x}_{\mathcal{V}_J}) = 1$; 
 it also remarked that the configuration $ \boldsymbol{x}_{\mathcal{V}_J}^{\star}$ denotes the elements of  $ \boldsymbol{x}^{\star}$
associated with factor $\mathsf{g}_J$ and satisfies $\mathsf{g}_J(\mathbf{x}_{\mathcal{V}_J}=
 \boldsymbol{x}_{\mathcal{V}_J}^{\star}) = 1$.
 Given that Theorem~\ref{theorem:convergence_valid_solution_every_n_natural} assumes existence of at least one solution, the configuration space of the summation in Eq.~\eqref{fac_to_var_iter1_bit0} contains at least
$1$ feasible configuration $\mathbf{x}_{\mathcal{V}_J}$  ($\boldsymbol{x}_{\mathcal{V}_J}^{\star}$
 is one of them).
 Thus, using the above observation and substituting
  Eq.~\eqref{eq:mu_var_to_fac_iter0} 
  in~\eqref{fac_to_var_iter1_bit0},  the
 following inequality is obtained: 
\begin{align}
 \mathsf{m}^{(1)}_{ J \rightarrow  {v_0}}\!\!\left( \mathtt{x}_{v_{0}}^{\star} \right)  &=  \mathtt{C}_{J \rightarrow 
 {v_0}}^{(1)}(\mathbf{q})  \sum_{\begin{smallmatrix} \mathbf{x}_{\mathcal{V}_J} :  \mathsf{g}_J(\mathbf{x}_{\mathcal{V}_J}) = 1 \\ x_{v_0} =
  \mathtt{x}_{v_{0}}^{\star}  \end{smallmatrix}} 
\prod_{y \in \mathcal{V}_J \backslash v_0}\!\!  \text{P}_{y}(x_y)  \nonumber   \\
&\geq   \mathtt{C}_{J \rightarrow  v_0}^{(1)}(\mathbf{q}) \! \! \prod_{y \in \mathcal{V}_J \backslash v_0 } \!\!\!  
\text{P}_{ y} \!\left( \mathtt{x}_{y}^{\star} \right). 
 \label{ineq_1}
\end{align}

The marginal of variable $x_{v_0}$
for the first iteration is given by Eq.~\eqref{marginal} for $n=1$;
using Eqs.~\eqref{def_prior} and~\eqref{ineq_1}, the marginal
for $  x_{v_0} = \mathtt{x}_{v_{0}}^{\star} $  is lower bounded:
\begin{align}
\mathsf{r}_{v_0}^{(1)}\!\left( \mathtt{x}_{v_{0}}^{\star} \right)  &=    \text{P}_{ v}( \mathtt{x}_{v_0}^{\star})  \prod_{J \in \mathcal{J}_{v_0} }  \mathsf{m}^{(1)}_{ J \rightarrow  v_0}( \mathtt{x}_{v_0}^{\star})  \nonumber\\
& \overset{  }{>}  ( 1-\epsilon_{v_0})     \prod_{J \in \mathcal{J}_{v_0} } 
\mathtt{C}_{J \rightarrow  v_0}^{(1)}(\mathbf{q}) \! \! \prod_{y \in \mathcal{V}_J \backslash v_0 } \!\!\!  
\text{P}_{ y} \!\left( \mathtt{x}_{y}^{\star} \right) .
\label{marginal_iter_1_for_bit0}
\end{align}

An upper bound for $\mathsf{r}_{v_0}^{(1)}\!\left(1- \mathtt{x}_{v_{0}}^{\star} \right) $ is found, exploiting Eq.~\eqref{def_prior}, i.e., $\text{P}_{ v}\!\left(  \mathtt{x}_{v}^{\star}\right) > \frac{1}{2} > \text{P}_{ v}\!\left(1-  \mathtt{x}_{v}^{\star}\right)$
 for any $v \in \mathcal{V}$. For any   $J \in \mathcal{J}_{v_0}$,  
 \begin{subequations}\label{eq:argmax_x_V_j_A}
\begin{align}
\prod_{y \in   \mathcal{V}_{J} \backslash v_0}  \text{P}_{ y}( \mathtt{x}_{y}^{\star}) 
&=   \max_{\mathbf{x}_{\mathcal{V}_J
\backslash v_0} \in \mathds{B}^{|\mathcal{V}_J|-1}} \!
\prod_{y \in   \mathcal{V}_{J} \backslash v_0}  \text{P}_{ y}(x_y)
\label{eq:argmax_x_V_j_A_second} \\
~&\overset{\eqref{eq:mu_var_to_fac_iter0} }{ = }   \max_{\mathbf{x}_{\mathcal{V}_J
\backslash v_0} \in \mathds{B}^{|\mathcal{V}_J|-1}} \!\!
\prod_{y \in   \mathcal{V}_{J} \backslash v_0}\mathsf{m}^{(0)}_{ y \rightarrow J}(x_y). 
\label{eq:argmax_x_V_j_A_first} 
\end{align}
 \end{subequations}
Using Eq.~\eqref{eq:mu_var_to_fac_iter0} in \eqref{fac_to_var_iter1_bit1}, applying the definition of $\kappa_{v_0}$  and exploiting Eq.~\eqref{eq:argmax_x_V_j_A},
\begin{align}
 \mathsf{m}^{(1)}_{ J \rightarrow  {v_0}}\!\left( 1- \mathtt{x}_{v_0}^{\star} \right)   \!&   = 
  \mathtt{C}_{J \rightarrow  v_0}^{(1)}(\mathbf{q})  \!\! \sum_{\begin{smallmatrix} \mathbf{x}_{\mathcal{V}_J} :  \mathsf{g}_J(\mathbf{x}_{\mathcal{V}_J}) = 1 \\ x_{v_0}=  1- \mathtt{x}_{v_0}^{\star} \end{smallmatrix}}  \prod_{y \in \mathcal{V}_J \backslash  v_0 }\!\! \text{P}_{ y}(x_y) \nonumber
\\
&
\overset{\eqref{eq:k1}}{ \leq } \mathtt{C}_{J \rightarrow  v_0}^{(1)}(\mathbf{q})~  \kappa_{v_0}\!\! \!  \max_{\mathbf{x}_{\mathcal{V}_J
\backslash v_0} \in \mathds{B}^{|\mathcal{V}_J|-1}} \!\! \!
\prod_{y \in   \mathcal{V}_{J} \backslash v_0} \!\!  \text{P}_{ y}(x_y)  \nonumber \\
&\overset{\eqref{eq:argmax_x_V_j_A}    }{=}
\mathtt{C}_{J \rightarrow  v_0}^{(1)}(\mathbf{q})~  \kappa_{v_0} \prod_{y \in   \mathcal{V}_{J} \backslash v_0}  \text{P}_{ y}( \mathtt{x}_{y}^{\star}) .
\label{eq:ineq_mu_J_v_0_iter_1}
\end{align}
Therefore,  the marginal of $v_0$ satisfies the following:
\begin{align}
\mathsf{r}_{v_0}^{(1)}\!\left( 1- \mathtt{x}_{v_0}^{\star} \right) &=   
\text{P}_{ v}(1- \mathtt{x}_{v_{0}}^{\star}) \prod_{J \in \mathcal{J}_{v_0} }
\mathsf{m}^{(1)}_{ J \rightarrow  v_0}\!\!\left( 1- \mathtt{x}_{v_0}^{\star} \right)  \nonumber \\
&
\overset{\eqref{def_prior}}{\underset{\eqref{eq:ineq_mu_J_v_0_iter_1} }{ <}}
  \epsilon_{v_0}      \prod_{J \in \mathcal{J}_{v_0} }  \kappa_{v_0} ~  \mathtt{C}_{J \rightarrow  {v_0}}^{(1)}(\mathbf{q}) 
 \prod_{y \in   \mathcal{V}_{J} \backslash v_0}  \text{P}_{ y}( \mathtt{x}_{y}^{\star})  \nonumber \\
&\overset{}{=}   \epsilon_{v_0}   ~(\kappa_{v_0})^{ \left|\mathcal{J}_{v_0} \right|}\!\!   \prod_{J \in \mathcal{J}_{v_0} }   \mathtt{C}_{J \rightarrow  {v_0}}^{(1)}(\mathbf{q})  \!\!
  \prod_{y \in   \mathcal{V}_{J} \backslash v_0}  \text{P}_{ y}( \mathtt{x}_{y}^{\star})
  \nonumber \\
&\overset{(a)}{=}  (1-\epsilon_{v_0})
   \prod_{J \in \mathcal{J}_{v_0} }   \mathtt{C}_{J \rightarrow  {v_0}}^{(1)}(\mathbf{q}) 
 \prod_{y \in   \mathcal{V}_{J} \backslash v_0}  \text{P}_{ y}( \mathtt{x}_{y}^{\star}) \nonumber  \\
&\overset{\eqref{marginal_iter_1_for_bit0}}{<} \mathsf{r}_{v_0}^{(1)}\!\left( \mathtt{x}_{v_{0}}^{\star} \right)  ,
\label{eq:marginal_comparison_iter1}
\end{align}
where  step $ (a)$ above used that $\epsilon_{v_0}\! \left((\kappa_{v_0})^{ \left|\mathcal{J}_{v_0} \right|} +1\right) = 1 $, stemming directly from the definition of $\epsilon_{v_0}$.
The choice of $v_0 \in \mathcal{V}$   and $J \in  \mathcal{J}_{v_0}$ was arbitrary  and thus, the proof is completed.
\end{IEEEproof}

\begin{lem} \label{lemma:m_x_v_g_j_one_iteration}
\normalfont
Under the assumptions of Theorem~\ref{theorem:convergence_valid_solution_every_n_natural}, 
\begin{equation}
\mathsf{m}^{(1)}_{ v \rightarrow  J}\!\left(  \mathtt{x}_{v}^{\star} \right)   >  \mathsf{m}^{(1)}_{ v \rightarrow  J}\!\left(1-  \mathtt{x}_{v}^{\star}\right ), \forall v\in \mathcal{V}, \forall J \in \mathcal{J}_v   .
\end{equation}
\end{lem}
\begin{IEEEproof} 
 Suppose that  prior  values $\{q_v\}_{v \in \mathcal{V}}$ satisfy Eq.~\eqref{priors_vals_sufficiency}.
Consider an arbitrary  $v_0 \in \mathcal{V}$ and $J \in \mathcal{J}_{v_0}$. Using the update rule in Eq.~\eqref{var_up_rule_1}
and applying the same reasoning with Lemma~\ref{lemma:convergence_valid_solution_one_iteration},
the following is offered:
\begin{align}
\frac{\mathsf{m}^{(1)}_{ v_0 \rightarrow  J}\!\left( \mathtt{x}_{v_0}^{\star} \right)}{\mathtt{C}_{v_0 \rightarrow  J}^{(1)}(\mathbf{q})} &= 
  \text{P}_{ v_0}\!\left( \mathtt{x}_{v_0}^{\star}\right)
 \prod_{I \in\mathcal{J}_{v_0}  \backslash  J} \mathsf{m}^{(1)}_{ {I} \rightarrow  v_0}\!\left( \mathtt{x}_{v_0}^{\star}\right)  \nonumber \\
&\overset{\eqref{def_prior}}{\underset{\eqref{ineq_1} }{ >}} (1-\epsilon_{v_0}) 
\!  \!  \prod_{I \in\mathcal{J}_{v_0}  \backslash  J}\! \! \mathtt{C}_{I \rightarrow  {v_0}}^{(1)}(\mathbf{q})\! \!\! \!
 \prod_{y \in   \mathcal{V}_{I} \backslash v_0}  \text{P}_{ y}( \mathtt{x}_{y}^{\star})   \nonumber \\
&=   \epsilon_{v_0}   ~(\kappa_{v_0})^{ \left|\mathcal{J}_{v_0} \right|}\! \!  \prod_{I \in\mathcal{J}_{v_0}  \backslash  J }\! \! \mathtt{C}_{I \rightarrow  {v_0}}^{(1)}(\mathbf{q})\! \!\! \!
 \prod_{y \in   \mathcal{V}_{I} \backslash v_0}  \text{P}_{ y}( \mathtt{x}_{y}^{\star}) \nonumber 
 \\
&\overset{(a)}{\geq} \! \epsilon_{v_0} \!  \! \!  \prod_{I \in\mathcal{J}_{v_0}  \backslash  J }\! \! \mathtt{C}_{I \rightarrow  {v_0}}^{(1)}(\mathbf{q})~\kappa_{v_0}\! \!\! \!
\prod_{y \in   \mathcal{V}_{I} \backslash v_0}  \text{P}_{ y}( \mathtt{x}_{y}^{\star})  \nonumber
\\
&\overset{\eqref{def_prior}}{\underset{\eqref{eq:ineq_mu_J_v_0_iter_1} }{ >}} \! \text{P}_{ v_0}\!\left(1- \mathtt{x}_{v_0}^{\star}\right)  \! \! 
 \prod_{I \in\mathcal{J}_{v_0}  \backslash  J} \mathsf{m}^{(1)}_{ {I} \rightarrow  v_0}\!\left(1- \mathtt{x}_{v_0}^{\star}\right) \nonumber
\\
&\overset{}{=} \frac{\mathsf{m}^{(1)}_{ v_0 \rightarrow  J}\!\left(1- \mathtt{x}_{v_0}^{\star}\right)}{\mathtt{C}_{v_0 \rightarrow  J}^{(1)}(\mathbf{q})},
\label{eq:ineq_Lemma2}
\end{align}
where step $ (a)$ above used that   $ (\kappa_{v_0})^{ \left|\mathcal{J}_{v_0} \right| }\geq 
(\kappa_{v_0})^{ \left|\mathcal{J}_{v_0} \right|-1}$, due to  the fact that $\kappa_{v_0} \geq 1$. The choice of $v_0 \in \mathcal{V}$ and $J \in  \mathcal{J}_{v_0}$ was arbitrary and thus, the proof is completed.
\end{IEEEproof}

\begin{IEEEproof}[Proof of Theorem~\ref{theorem:convergence_valid_solution_every_n_natural}]
Suppose that Eq.~\eqref{priors_vals_sufficiency} holds. The theorem will be proved by induction. 

Denote for all $n \in \mathds{N} \cup \{0\}$,
\begin{equation}
\mathfrak{T}_{\mathsf{m}}^{(n)}  \equiv  \prod_{ v \in \mathcal{V}} 
 \prod_{ J \in \mathcal{J}_v }   
 \mathsf{1}\!\left\{ \mathsf{m}^{(n)}_{ v \rightarrow  J}\!\left( \mathtt{x}_{v}^{\star} \right)   >  \mathsf{m}^{(n)}_{ v \rightarrow  J}\!\left(1- \mathtt{x}_{v}^{\star}\right ) \right \}  \label{eq:statement_mu_v_J_n}
\end{equation}
and for all $n \in \mathds{N}$,
\begin{equation}
\mathfrak{T}_{\mathsf{r}}^{(n)}  \equiv  \prod_{ v \in \mathcal{V}} 
\mathsf{1}\!\left\{
  \mathsf{r}^{(n)}_{ v}\!\left( \mathtt{x}_{v}^{\star} \right)   >  \mathsf{r}^{(n)}_{ v }\!\left(1- \mathtt{x}_{v}^{\star}\right ) \right \}.
 \label{eq:statement_rho_v_n}
\end{equation}
Initialization according to Eq.~\eqref{priors_vals_sufficiency} satisfy Eq.~\eqref{eq:mu_var_to_fac_iter0}, which is equivalent to $\mathfrak{T}_{\mathsf{m}}^{(0)}= 1$. Such condition satisfaction offers $ \mathfrak{T}_{\mathsf{r}}^{(1)}$
according to Lemma~\ref{lemma:convergence_valid_solution_one_iteration} and $ \mathfrak{T}_{\mathsf{m}}^{(1)}$ according to Lemma~\ref{lemma:m_x_v_g_j_one_iteration}. Therefore, the following holds:
\begin{equation}
\mathfrak{T}_{\mathsf{m}}^{(0)}= 1 \overset{\eqref{priors_vals_sufficiency}}{\underset{ }{ \implies}} \mathfrak{T}_{\mathsf{r}}^{(1)}=1 \text{ and }  \mathfrak{T}_{\mathsf{m}}^{(1)} = 1.
\end{equation}
Subsequent section establishes the following: 
\begin{equation}
\mathfrak{T}_{\mathsf{m}}^{(n)} =1 \overset{\eqref{priors_vals_sufficiency}}{\underset{ }{ \implies}} 
 \mathfrak{T}_{\mathsf{r}}^{(n+1)}=1 \text{ and }  \mathfrak{T}_{\mathsf{m}}^{(n+1)}=1,
\label{eq:indunction_step}
\end{equation}
implying that Eq.~\eqref{eq:marginal_final_forall_n} is true. 

Assume that the induction hypothesis holds, i.e.,  $\mathfrak{T}_{\mathsf{m}}^{(n)}=1$. We now show the right-hand side of Eq.~\eqref{eq:indunction_step}. Choosing an arbitrary  $v_0 \in \mathcal{V}$,
similarly to~\eqref{ineq_1},
for any    $J \in \mathcal{J}_{v_0}$, the message $\mathsf{m}^{(n+1)}_{ J \rightarrow  {v_0}}\!\left( \mathtt{x}_{v_0}^{\star}\right) $
can be upper bounded as follows:
\begin{equation}
 \mathsf{m}^{(n+1)}_{ J \rightarrow  {v_0}}\!\left( \mathtt{x}_{v_0}^{\star}\right)    \geq
   \mathtt{C}_{J \rightarrow  v_0}^{(n+1)}(\mathbf{q})  \! \!  \prod_{y \in \mathcal{V}_J \backslash v_0}
 \! \mathsf{m}^{(n)}_{ y \rightarrow  J}\left( \mathtt{x}_{y}^{\star} \right).
\label{eq:ineq_mu_J_v_0_iter_n_plus_1} 
\end{equation}
Using the induction hypothesis   $\mathfrak{T}_{\mathsf{m}}^{(n)}=1$, which   implies that $\mathsf{m}^{(n)}_{ v \rightarrow  J}\!\left( \mathtt{x}_{v}^{\star} \right) 
>  \mathsf{m}^{(n)}_{ v \rightarrow  J}\!\left(1- \mathtt{x}_{v}^{\star}\right )$, $\forall  v \in \mathcal{V} ,
 J \in \mathcal{J}_v    $,   under the same reasoning followed in~\eqref{eq:argmax_x_V_j_A_first}, the following is obtained:
\begin{equation}
\prod_{y \in   \mathcal{V}_{J} \backslash v_0} \! \mathsf{m}^{(n)}_{ y \rightarrow J}( \mathtt{x}_{y}^{\star})  =  \! \max_{\mathbf{x}_{\mathcal{V}_J
\backslash v_0} \in \mathds{B}^{|\mathcal{V}_J|-1}} \! \!
\prod_{y \in   \mathcal{V}_{J} \backslash v_0} \! \mathsf{m}^{(n)}_{ y \rightarrow J}(x_y ).
\label{eq:argmax_x_V_j_A_n_th_step} 
\end{equation}
Hence, working as in Eq.~\eqref{eq:ineq_mu_J_v_0_iter_1} and using Eq.~\eqref{eq:argmax_x_V_j_A_n_th_step},  for any    $J \in \mathcal{J}_{v_0}$, the message $\mathsf{m}^{(n+1)}_{ J \rightarrow  {v_0}}\!\left(1- \mathtt{x}_{v_0}^{\star}\right) $ is upper bounded:
\begin{equation}
\mathsf{m}^{(n+1)}_{ J \rightarrow  {v_0}}\left(1- \mathtt{x}_{v_0}^{\star} \right) \leq
\mathtt{C}_{J \rightarrow  v_0}^{(n+1)}(\mathbf{q})~  \kappa_{v_0} \!\! \prod_{y \in \mathcal{V}_J \backslash v_0} \!   \mathsf{m}^{(n)}_{ y \rightarrow  J}\left( \mathtt{x}_{y}^{\star}\right).
\label{eq:ineq_mu_J_v_1_iter_n_plus_1}
\end{equation}
Substituting Eqs.~\eqref{eq:ineq_mu_J_v_0_iter_n_plus_1} and~\eqref{eq:ineq_mu_J_v_1_iter_n_plus_1} 
  in $\mathsf{r}_{v_0}^{(n+1)}\!\left( \mathtt{x}_{v_0}^{\star}\right) $ and
$\mathsf{r}_{v_0}^{(n+1)}\!\left(1- \mathtt{x}_{v_0}^{\star}\right) $, respectively,  
after similar algebra as in Eq.~\eqref{eq:marginal_comparison_iter1}, the following is obtained:
\begin{equation}
\frac{\mathsf{r}_{v_0}^{(n+1)}\!\left( \mathtt{x}_{v_0}^{\star}\right)  }{  \mathsf{r}_{v_0}^{(n+1)}\!\left(
1- \mathtt{x}_{v_0}^{\star}\right)} 
\overset{ }{>} \frac{1-\epsilon_{v_0}}{\epsilon_{v_0} }\frac{1}{( \kappa_{v_0})^{|\mathcal{J}_{v_0}|}} 
= 1 .\label{eq:marginals_geq_1_forall_n}
\end{equation}
In a similar vein as above, using the same reasoning as in Eq.~\eqref{eq:ineq_Lemma2}, the following is obtained: 
\begin{equation}
\frac{\mathsf{m}^{(n+1)}_{ v_0 \rightarrow  J}\!\left( \mathtt{x}_{v_0}^{\star} \right)}{\mathsf{m}^{(n+1)}_{ v_0 \rightarrow  J}\!\left(1- \mathtt{x}_{v_0}^{\star} \right)}
\overset{ }{>}
\frac{1-\epsilon_{v_0}}{\epsilon_{v_0} }\frac{1}{ ( \kappa_{v_0})^{\left(|\mathcal{J}_{v_0}|-1\right)}} 
\geq
1.
\end{equation}

The  choice of $v_0 \in \mathcal{V}$ and $J \in  \mathcal{J}_{v_0}$ was arbitrary and thus, 
the induction step in Eq.~\eqref{eq:indunction_step} is established, 
proving the theorem. Extension to the damped version can be obtained similarly.
\end{IEEEproof}

\balance

\bibliographystyle{IEEEtran}
\bibliography{IEEEabrv,FrecAlloc_v16}

\end{document}